\documentclass[iop]{emulateapj-rtx4} 
\shortauthors{Sekanina}
\shorttitle{New Model for Kreutz Sungrazer System}
\slugcomment{Version \today }
\newcommand{\Rsun}{$R_{\mbox{\scriptsize \boldmath $\odot$}}\!$}

\begin{document}
\title{NEW MODEL FOR THE KREUTZ SUNGRAZER SYSTEM:\ CONTACT-BINARY PARENT AND\\
       UPGRADED CLASSIFICATION OF DISCRETE FRAGMENT POPULATIONS}
\author{Zdenek Sekanina}
\affil{Jet Propulsion Laboratory, California Institute of Technology,
  4800 Oak Grove Drive, Pasadena, CA 91109, U.S.A.}
\email{Zdenek.Sekanina@jpl.nasa.gov.}

\begin{abstract} 
The structure of the Kreutz system of sungrazing comets is shown to be much more complex than
formerly believed.  Marsden's (1989) division into three subgroups (I, II, IIa) is now greatly
expanded, as new evidence is being offered on five main populations of fragments --- I, Ia, II,
IIa, and III --- and four peripheral ones --- Pre-I, Pe (side branch of I), IIIa, and IV.  Five
populations are related to naked-eye sungrazers and all nine incorporate carefully screened data
sets from a collection of 1500~SOHO/STEREO dwarf Kreutz comets whose gravitational orbits were
computed by Marsden.  Tight correlations between the nominal perihelion latitude and the nominal
longitude of the ascending node are the result of ignored effects of an outgassing-driven
acceleration on the orbital motion.  The average width of a gap between adjacent populations in
the nodal longitude corrected for the nongravitational effect is near 9$^\circ$; the overall
range equals 66$^\circ$.  The other angular elements of the populations are determined by the
condition of shared apsidal line.  A self-consistent model postulates (i)~an initial breakup,
in general proximity of aphelion, of a contact-binary parent (progenitor) into its two lobes
and the neck (originally linking the lobes), giving birth to, respectively, Population~I
(Lobe~I; the main residual mass C/1843~D1), Population~II (Lobe~II; C/1882~R1), and
Population~Ia (the neck); followed by (ii)~progressive fragmentation of the lobes (primarily
Lobe~II), mostly (but not exclusively) far from perihelion, giving successively rise to the
other populations and clusters of naked-eye Kreutz sungrazers and their debris.  The separation
velocities were a few meters per second.  Massive fragments of Populations Pre-I, IIIa, and IV
are yet to be discovered.  Relations among the products of cascading fragmentation are depicted
in a pedigree chart.  The age of the Kreutz system is estimated at two millennia and a mean
orbital period of Lobe~I and its main residual mass at $\sim$740~yr.  The status is reviewed of
the possible historical Kreutz comets seen~in~AD~1106,~AD~363,~and~372~BC.
%
\end{abstract}

\section{Introduction} 
\vspace{0.2cm}
This paper is the preliminary report of a comprehensive investigation, to follow shortly, aimed
at introducing a new fragmentation model for the Kreutz sungrazer system --- a {\it major
revision\/} of{\vspace{-0.07cm}} the diverse orbital work conducted over the past 60~years by,
among others, \"{O}pik (1966), Marsden (1967, 1989, 2005), Seka\-nina (2002), and, in particular,
Sekanina \& Chodas (2004, 2007) --- the two studies that employed competing rationales.
Necessitated by a number of key developments during the past decade, the new model accounts
for and/or is consistent with:\ (i)~the appearance of the sungrazer C/2011~W3 (Lovejoy), an
event of singular importance for revising the orbital classification of the Kreutz system;
(ii)~a nearly steady stepwise distribution of the longitudes of the ascending node among the
naked-eye sungrazers and a correlation with the perihelion distance; (iii)~growing evidence
that the contact binary, ostensibly susceptible to breakup, is a fairly common figure among
cometary nuclei and Kuiper Belt objects; (iv)~a proposition that C/1843~D1 and C/1882~R1
are the two dominant surviving residual masses of the progenitor, the nearly 40~year wide
gap between their perihelion times being the outcome of the initial rate of separation of
their precursors, magnified by the orbital evolution over the lifetime of the Kreutz system;
(v)~a novel look at some candidates for the Kreutz-system membership among historical comets;
and (vi)~results from my examination of Marsden's collection of gravitational orbits for a
large number of dwarf Kreutz sungrazers detected by the coronagraphs on board the Solar and
Heliospheric Observatory (SOHO) and the Solar Terrestrial Relations Observatory (STEREO).

With regard to point (vi), one should remark that the gravitational orbit approximates the
motion of a Kreutz sungrazer the better the more massive the object is.  While the deviation
from the gravitational law is trivial and essentially undetectable in the motion of a
brilliant, naked-eye member of the Kreutz system, the motion of a faint dwarf sungrazer could
defy the law to a considerable degree, as argued in Section~3.  Accordingly, the computation
of a gravitational orbit makes sense for bright sungrazers, for which it offers an excellent
fit to astrometric observations, but it is dynamically meaningless for many dwarf comets.
Because the observed orbital arc of a dwarf sungrazer is short and, as pointed out below, its
astrometry is inferior, the gravitational fit may look acceptable.  Yet, the derived orbit is
spurious, the angular elements often grossly failing to satisfy the well-known condition of
shared apsidal line.\footnote{As far as I am aware, the coincidence of the apsidal lines of
C/1843~D1 and C/1882~R1 was discovered by Kreutz (1895).  He was greatly impressed by the
match, which contrasted with the large differences in the nodal longitudes and other angular
elements of the two comets.  The general validity of the condition of shared apsidal line
among the Kreutz sungrazers with well-determined orbits was underscored by Marsden (1967).}
Nonetheless, I demonstrate in Section~3 that when properly examined, Marsden's gravitational
orbits provide important information on major features of the orbital distribution of the
Kreutz system's dwarf comets.

\section{Upgrading Marsden's Classification:\ From Subgroups to Populations} 
When Marsden (1967) was writing his first paper on the subject, fewer than ten Kreutz
sungrazers were known with moderately to highly accurate orbits.  He referred to their total
as the Kreutz {\it group\/} and, based on their orbital diversity, he divided them into two
{\it subgroups\/}, I and II.  Given the thousands of members known nowadays, I deem it more
appropriate to call their total the Kreutz {\it system\/} and to divide the ensemble into
fragment {\it populations\/}.  Accordingly, in the following I~consistently replace the term
used by Marsden with the new one, even when I refer to his own work:\ his Subgroups~I and II
are now Populations~I and II, respectively. 

Of the sungrazers known in the 1960s, Marsden (1967) rated C/1843~D1 (then designated 1843~I),
C/1880~C1 (1880~I), and C/1963~R1 (1963~V) as definite members of Population~I; whereas
C/1882~R1 (1882~II), C/1945~X1 (1945~VII), and{\vspace{-0.04cm}} C/1965~S1 (1965~VIII) as definite
members of Population~II.\footnote{Marsden's assigning C/1945~X1, the only telescopic comet on his
list, to Population~II was rather problematic (Marsden 2005).  Even nowadays the orbit is
deemed too uncertain to fit unequivocally any particular population.}  Of the other comets,
including several historical ones, Marsden regarded C/1668~E1 (Southern of 1668), C/1695~U1
(Southern of 1695), X/1882~K1 (Tewfik; eclipse comet of 1882), and C/1887~B1 (1887~I) as
potential{\vspace{-0.04cm}} members of Population~I,\footnote{The orbit computed for C/1887~B1 by
Sekanina (Marsden \& Roemer 1978), nowadays accepted as the most representative and making the
comet's suspected membership in Population~I more secure, was unavailable at the time of Marsden's
(1967) paper.} while he believed that C/1689~X1 (Southern of 1689) and X/1702~D1 (1702a)
were likely to belong to Population~II.

The orbital differences between Populations I and II are by no means minor; they amount to
up to about 20$^\circ$ in the angular elements and approximately 0.6~{\Rsun} (or up to some
50~percent) in the perihelion distance.  Specifically, the definite members of Population~I
have the longitudes of the ascending node, $\Omega$, between 3$^\circ$ and 8$^\circ$ and
the perihelion distances, $q$, not exceeding 1.2~{\Rsun}\,, while the definite members of
Population~II have $\Omega$ near 347$^\circ$ and $q$ near 1.7~{\Rsun}\,.  The members of both
populations satisfy the condition of shared apsidal line, defined by the {\it standard\/}
perihelion longitude \mbox{$L_\pi = 282^\circ\!.8 \pm 0^\circ\!.2$} and the {\it standard\/}
perihelion latitude \mbox{$B_\pi = +35^\circ\!.2 \pm 0^\circ\!.1$}.\footnote{All angular
elements in this paper refer to equinox J2000.}  The longitude of the ascending node is shown
to be the orbital element that most faithfully describes the population to which a given
sungrazer belongs.  The remaining angular elements --- the argument of perihelion, $\omega$,
and the inclination, $i$ --- are fully determined by $\Omega$, $L_\pi$, and $B_\pi$.  The
perihelion distance does not appear to be population diagnostic to the degree the nodal
longitude is.   

The sungrazer C/1970 K1, arriving nearly three years after Marsden's pioneering paper had
been published, failed to fit either of the two populations in terms of both the angular
elements and the perihelion distance.  To account for the anomaly, Marsden (1989) expanded
his classification system by introducing Population~IIa.  With this comet's value of $\Omega$
at 337$^\circ$, the range of nodal longitudes was extended from 20$^\circ$ to 30$^\circ$; and
with its $q$ near 1.9~{\Rsun}\,, the range of perihelion distances increased to more than
0.8~{\Rsun}\,.  Good news was that C/1970~K1 did satisfy the condition of shared apsidal line.

Only one year after Marsden's untimely death, T.\,Lovejoy discovered comet C/2011~W3, which
failed to fit the three-population system.  With the longitude of the ascending node near
327$^\circ$, this comet extended the range of $\Omega$ among the naked-eye Kreutz sungrazers
by yet another $\sim$10$^\circ$ to $\sim$40$^\circ$.  Although this comet did fit, perhaps
fortuitously, Population~I in terms of the perihelion distance, further expansion of the
classification system was unavoidable;{\vspace{-0.05cm}} comet C/2011~W3 became a representative
of a new Population~III.\footnote{This very suggestion was made in another context by Sekanina \&
Chodas (2012).}

At this point in time there were four populations, I, II, IIa, and III, with a peculiar nodal-longitude
distribution:\ the gap between Populations~I and II was $\sim$20$^\circ$ wide, but the gaps
between Populations~II and IIa and between Populations~IIa and III were each $\sim$10$^\circ$
wide.  For the sake of uniformity, one is tempted to speculate that yet another, hidden
population --- I call it Population~Ia --- resides near \mbox{$\Omega \simeq 356^\circ$},
approximately halfway between Populations~I and II.  If so, the gap widths between adjacent
populations become equalized at about 9$^\circ$ to 10$^\circ$ over the entire range of nodal
longitudes between Populations~I and III.

Circumstantial evidence of Population~Ia is offered by the orbital distribution for a group
of eight SOHO Kreutz sungrazers, corrected by Sekanina \& Kracht (2015) for {\it major
nongravitational effects\/} (Section~3).  The corrected longitudes of the ascending node for
three of the eight comets range from 358$^\circ$ to 1$^\circ$, outside the interval occupied
by the naked-eye sungrazers of Population~I, but near the above proposed nodal longitude of
Population~Ia (Table~1). 

Another set of Kreutz sungrazers worth examining for evidence of Population~Ia is 19 of the 20
comets imaged by the coronagraphs on board the Solwind (P78-1) and Solar Maximum Mission (SMM)
satellites between 1979 and 1989.\footnote{One of the 20~comets had a perihelion distance much
too large to qualify as a Kreutz sungrazer.}  Even though the orbits for these objects (see
Marsden 2005 for a review) are very uncertain, more so than the orbits of the SOHO sungrazers
of comparable brightness, and were, with a few exceptions, derived by forcing the condition of
shared apsidal line, at least three and possibly five among the nineteen appear to have moved
in orbits consistent with Population~Ia.\footnote{The likely Population~Ia members were
C/1981~V1 (Solwind~4), C/1983~S2 (Solwind~6), and C/1984~Q1 (Solwind~9); a possible member
was C/1981~W1 (Solwind~7) and perhaps even C/1987~T2 (SMM~1).}  In the following I show that
they represent the proverbial ``tip of the iceberg'' and that Population~Ia is far from being
the only contender in the search for additional populations among the dwarf Kreutz sungrazers.

\section{Marsden's Gravitational Orbits for the SOHO Kreutz Sungrazers} 
Marsden computed sets of gravitational orbital elements (i.e., with no nongravitational
terms in the equations of motion; the elements are referred to below as {\it nominal\/}) for
more than{\vspace{-0.03cm}} 1500~Kreutz sungrazers detected in the SOHO and STEREO coronagraphic
images.\footnote{Most of these orbits are listed in the Catalogue of Cometary Orbits
(Marsden \& Williams 2008), the rest is available in a number of the \mbox{2008--2010}
{\it Minor Planet Circular\/} issues.}  A plot of the nominal inclination $i$ against the
nominal longitude of the ascending node $\Omega$ for these dwarf Kreutz comets was used by
Sekanina \& Kracht (2015) to illustrate that the orbits failed to obey the condition of
shared apsidal line.  Another plot showed that the nominal latitude of perihelion $B_\pi$
(but not the longitude $L_\pi$) {\it varied systematically\/} with the nominal nodal
longitude $\Omega$.  The correlation disappeared and compliance with the condition of
shared apsidal line was restored when the orbital solution included the out-of-plane
component of the nongravitational acceleration.  The magnitude of the needed acceleration
--- assumed to be driven by the sublimation of water ice --- was enormous, ranging in
absolute value from 0.5 to \mbox{$25 \times\! 10^{-5}$\,AU day$^{-2}$} at 1~AU from the
Sun, or from 1.7~percent to 85~percent (sic) of the Sun's gravitational acceleration!  The
range of nodal longitudes for the eight dwarf sungrazers dropped from the nominal $\sim$90$^\circ$
to 11$^\circ$ after the nongravitational effect was accounted for.

Sekanina \& Kracht (2015) noted that in the plot of the nominal perihelion latitude against
the nominal longitude of the ascending node, most of the 1500 SOHO/STEREO Kreutz sungrazers
were located in a crowded branch along an approximately straight line with a slope of
\mbox{$dB_\pi/d\Omega = +0.28$}.  In addition, two parallel but sparsely populated
branches were apparent, passing through the positions that closely matched those of the
naked-eye sungrazers C/1970~K1 and C/2011~W3, respectively.  Because of large scatter, this
crude evidence on three populations of dwarf sungrazers was all that could be extracted
from the plot.

The low accuracy of Marsden's nominal orbits for the 1500 dwarf Kreutz sungrazers was a result
of the poor-quality astrometry of the coronagraphic images taken from aboard SOHO.\footnote{I
disregard below a small number of sungrazers whose orbits were based in part on images taken
with the coronagraphs on board the two STEREO spacecraft.}  However, given that astrometric
positions derived from images taken with the C2 coronagraph are a factor of about five more
accurate than positions from images taken with the C3 coronagraph, elimination from the set
of all sungrazers whose orbits relied in full or in part on the C3 positional data should
markedly improve the quality of the select subset of C2 based orbits, the curtailed length
of the orbital arc notwithstanding.  Removing, in addition, orbits based on too few
C2~positions,\footnote{The adopted rejection limit varied from population to population,
being stricter for populations with a greater number of members, and vice versa; see the row
$n_{\rm min}$ in Table~1 below.}  I ended up with 193~dwarf Kreutz sungrazers from the years
1997--2010, imaged exclusively with the C2 coronagraph.
%
\begin{figure*}[ht]
\vspace{-0.4cm}
\hspace{-.25cm}
\centerline{
\scalebox{0.79}{
\includegraphics{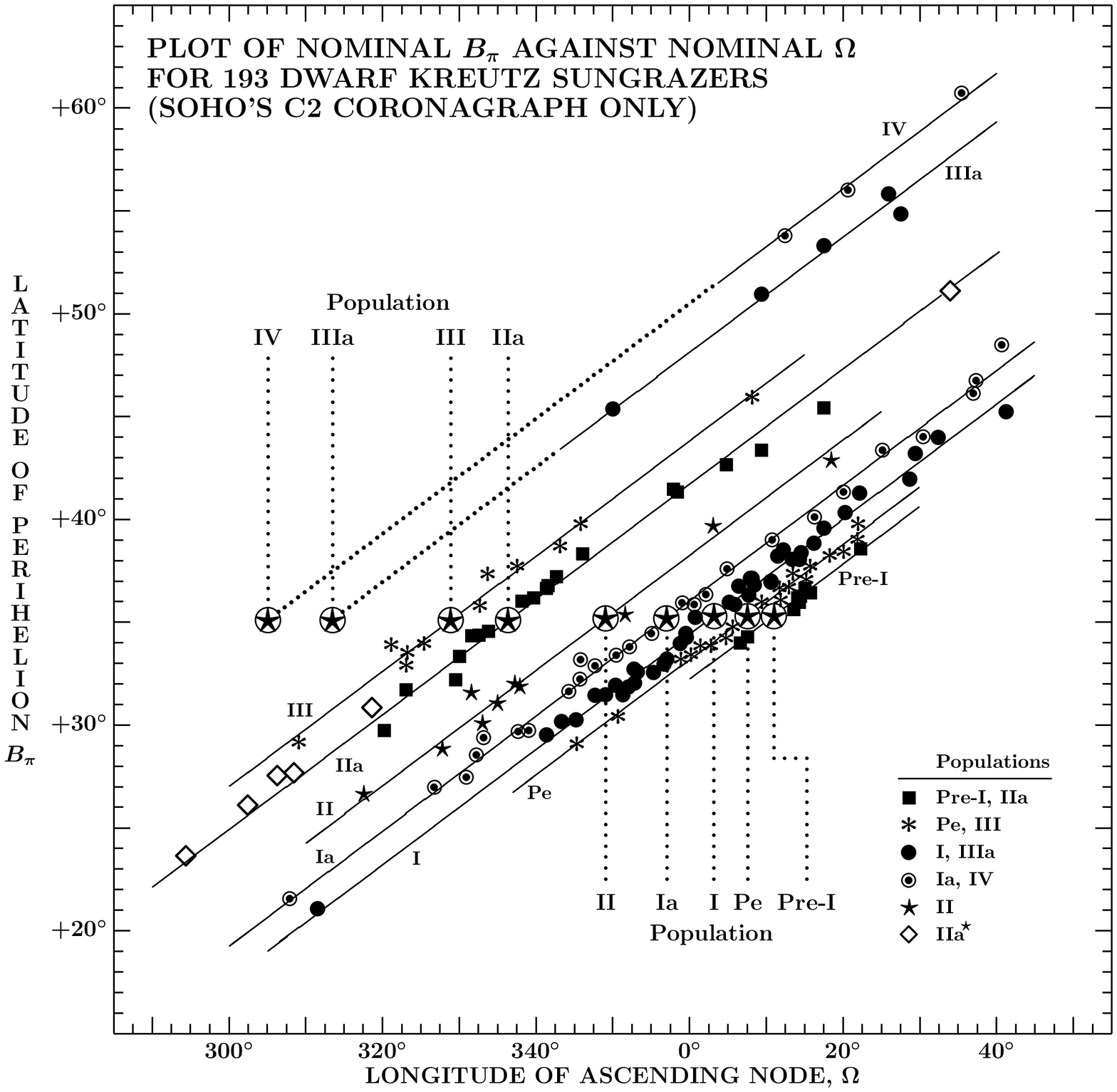}}}
\vspace{-7.25cm}
\caption{Plot of the nominal latitude of perihelion, $B_\pi$, as a function of the
nominal longitude of the ascending node, $\Omega$ (equinox J2000), for 193 dwarf
Kreutz sungrazers imaged in 1997--2010 exclusively with the C2 coronagraph on board
the SOHO Space Observatory; their gravitational orbits were computed by Marsden.  The
data points cluster fairly tightly along a set of straight lines of a constant slope
of \mbox{$dB_\pi/d\Omega = +0.28$} and refer to one of nine sungrazer populations.
Each line crosses the standard perihelion latitude of the population (Table~1) at
a point whose abscissa determines the respective population's true nodal longitude
(i.e., corrected for effects of the normal component of the nongravitational
acceleration).  In the order of decreasing true nodal longitude, the populations are
Pre-I, Pe, I, Ia, II, IIa, III, IIIa, and IV, as depicted in the plot; Population~Pe
is a side branch of Population~I.  All{\vspace{-0.08cm}} members of a population are plotted
with the same symbol; the exception is Population IIa, whose members with anomalously small
perihelion distances, IIa{\large \boldmath $^{\!\star}\!$} (Section~5.2), are identified by
symbols that differ from the symbols for the remaining members.  The position of the
{\it true\/} longitude of the ascending node is for each population highlighted by an
oversized circled star.  Some data points, which were overlapping any of these major
symbols or were contributing to one of awkward-looking local clumps, were either
removed or slightly shifted along the slope of the fitting lines.{\vspace{0.5cm}}}
\end{figure*}

The plot of a nominal $B_\pi$ against a nominal $\Omega$ for this carefully screened set of
sungrazers in Figure~1 confirms the universal validity of the constant slope $dB_\pi/d\Omega$
of +0.28, derived now with an estimated error of less than $\pm$2~percent.  Astonishingly,
the plot shows that the {\it number of dwarf-sungrazer populations\/} revealed by this set of
select data is {\it substantially higher\/} than the number seen in a plot of the unrestricted
set of 1500~SOHO/STEREO objects, thanks unquestionably to the much better data quality.

The comets in the crowded branch, previously assigned to Population~I, turn out instead to
be distributed among {\it four\/} populations.  Next to Population~I itself, which extends
over a 90$^\circ$ wide range of nominal nodal longitudes, contains nearly 50~percent of the
data, and appears to be closely linked to comet C/1843~D1, this branch also contains newly
recognized sets of dwarf comets:\ (i)~Population~Pre-I, spanning a range of only 16$^\circ$
in the nominal nodal longitude, is centered on \mbox{$\Omega \simeq 15^\circ$} (and may have
been detected among the SMM and Solwind sungrazers); (ii)~Population~Pe, clearly related to
comet C/1963~R1 (Pereyra) and designated by the initial letters of the discoverer's name,
extends over a range of nominal nodal longitudes from 345$^\circ$ to 22$^\circ$; and (iii)~the
already suspected Population~Ia, extending in the nodal longitude as widely as Population~I,
discriminates from the latter quite well over some intervals of $\Omega$, but less well over
others.  Because C/1963~R1 is a member of Population~I, the associated Population~Pe is merely
a side branch of Population~I.  Population~Pre-I, which was never before suspected to exist
and whose naked-eye sungrazer is still awaiting discovery, is likely to have separated from
Population~I or Pe at some point in the past.

Dwarf sungrazers of Populations~II (the main comet C/1882~R1), IIa (C/1970~K1), and III
(C/2011~W3) are clearly seen in Figure~1, the widest range of nominal nodal longitudes being
exhibited by IIa.  Major surprise is the detection, beyond Population~III, of two additional,
though apparently minor, Populations~IIIa and IV.  No naked-eye sungrazers are known as yet,
similarly to Population~Pre-I.  From the available data set there is no credible evidence of
the existence of additional fringe populations beyond the range from Pre-I to IV, at either
end of the range of corrected nodal longitudes (\mbox{$\Omega \gg 11^\circ$} or \mbox{$\Omega
\ll 305^\circ$}).
%
%

\begin{table*}[ht]
\vspace{-4.2cm}
\hspace{0.5cm}
\centerline{
\scalebox{1}{
\includegraphics{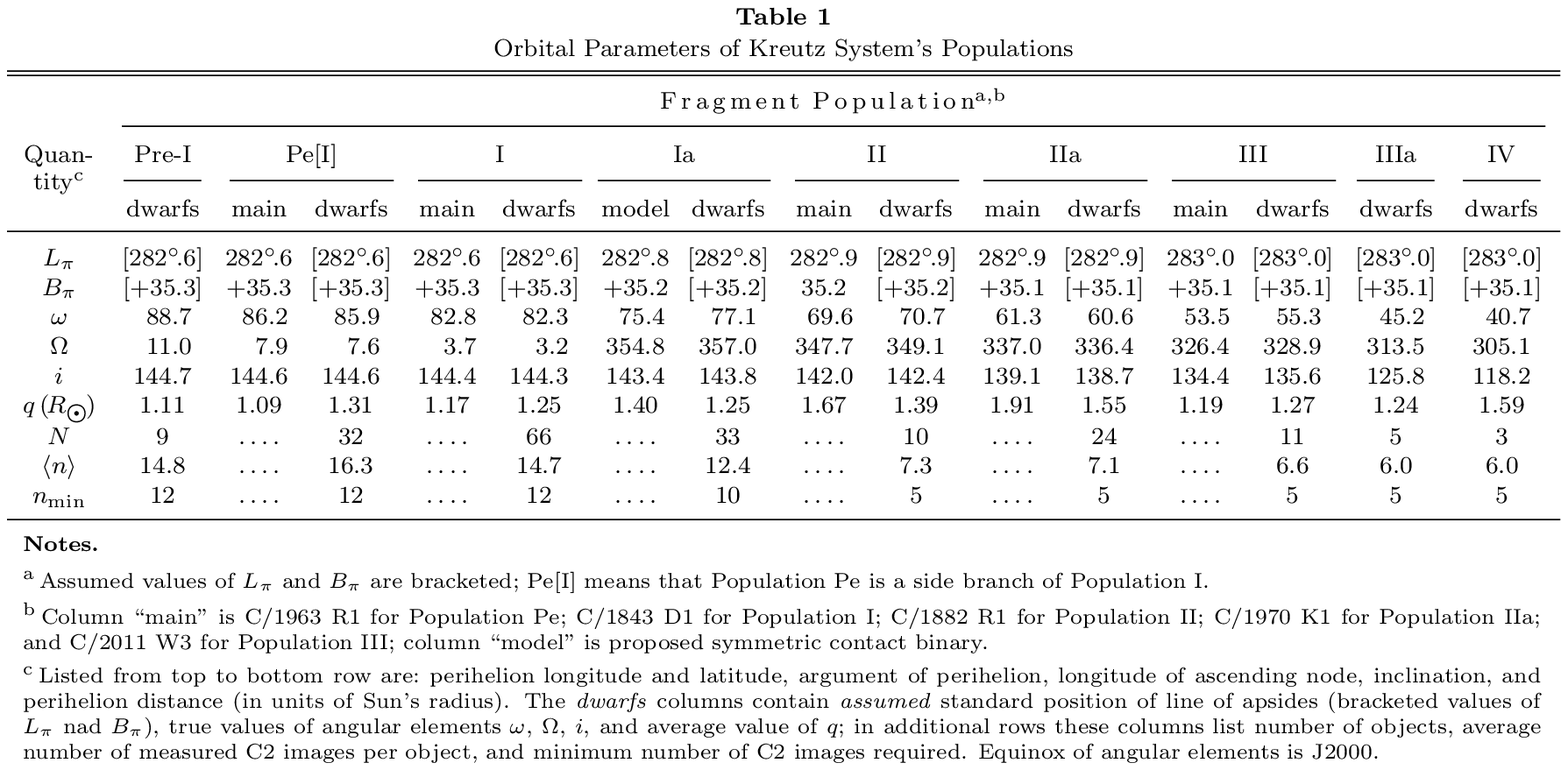}}}
\vspace{-16.3cm}
\end{table*}

The averaged orbits of the populations identified among the 193 select data are presented in
Table~1.  The first two rows offer the computed or adopted (when bracketed) {\it standard\/}
ecliptic coordinates of the direction to perihelion, the longitude $L_\pi$ and latitude
$B_\pi$, which are essentially constant.  Row~4 of the {\it dwarfs\/} columns lists for
each population the {\it true\/} longitude of the ascending node $\Omega$ (i.e., corrected
for effects of the out-of-plane component of the nongravitational acceleration), which is
given by the abscissa of the point on the line fitting the data in Figure~1, whose ordinate
equals the standard value of $B_\pi$ and which is marked by an oversized circled star.  This
true nodal longitude is then used, together with the standard values of $L_\pi$ and $B_\pi$,
to compute the true values of the other two angular elements, the argument of perihelion,
$\omega$, and the inclination, $i$, which are tabulated, respectively, in rows~3 and 5 of
the {\it dwarfs\/} columns of Table~1.  The average perihelion distance, $q$, is derived as
a mean of the perihelion distances of individual members of each population, which could
not be corrected for unknown effects of the transverse component of the nongravitational
acceleration.  Presented in row~6 of the {\it dwarfs\/} columns, these values of $q$ are
not population diagnostic, unlike the true nodal-longitude values.  With a single exception
(see below), this issue is not examined, but it may warrant a separate investigation in the
future.  The number of dwarf sungrazers, $N$, the average number of astrometric observations
used per sungrazer, $\langle n \rangle$, and the minimum number of observations required,
$n_{\rm min}$, are shown for each population in the last three rows of Table~1.

The magnitude of the nongravitational-acceleration's out-of-orbit component affecting the
motion of a dwarf sungrazer is approximately proportional to the distance of the nominal nodal
longitude from the true nodal longitude:\ the larger the distance, the greater the effect.
Figure~1 shows that the members of Populations~Pre-I and Pe were subjected to relatively low
accelerations, implying that the fragments were near an upper end of the size distribution
(perhaps many dozens of meters across); the members of Populations~I, Ia, and IIa were
subjected to a wide range of accelerations, implying a broad distribution of sizes; and
the members of Populations~IIIa and IV were exposed to very high accelerations, implying that
these fragments were near a lower end of the size distribution (possibly just meters across).

When one or more naked-eye sungrazers are known to be associated with a given population,
the orbit of the most prominent one is presented in column {\it main\/} of Table~1.  For
Population~Ia the main object is represented by the {\it model\/}, as described in Section~5.
Because of perturbations as well as uncertainties resulting from astrometric errors, the
{\it dwarfs\/} orbit differs from the {\it main\/} orbit of the same population, although
much less than are the differences in both the {\it dwarfs\/} and {\it main\/} orbits between
populations.  Table~1 shows that the largest {\it dwarfs\/}-to-{\it main\/} differences in
the nodal longitude, the element of premier interest, occur for Populations~III and Ia (in
excess of 2$^\circ$), while the smallest ones, less than 1$^\circ$, are for Populations~Pe,
I, and IIa.  The difference for Population~Ia might in part be due to the fact that the
assumptions affecting the {\it model\/} orbit, such as the lobe and neck symmetries
(Section~5), were unlikely to be satisfied.{\vspace{-0.1cm}}

\section{Orbital Effects of Fragmentation Events:\ Populations and Clusters} 
When a cometary nucleus splits into, say, two parts, either fragment leaves the point of
breakup with an orbital momentum that differs from the parent's orbital momentum.  This
difference is manifested by a separation velocity, whose vector sum with the parent's
orbital velocity at the point of breakup determines the fragment's heliocentric orbit
that deviates from the parent's.  In other words, the fragment has been perturbed, each
of its orbital elements by an amount that depends both on the magnitude and direction of
the separation velocity vector and, substantially, on the fragmentation event's orbital
location, which is entirely unrestricted. 

It turns out that perturbations of the inclination and the nodal longitude vary as the
distance from the Sun at fragmentation, whereas perturbations of the orbital period induced
by the transverse component of the separation velocity vary as the {\it inverse\/} distance
at fragmentation.\footnote{The orbital period is also perturbed by the radial component of
the separation velocity; this perturbation depends on the true anomaly, rather than the
distance, at the time of fragmentation.}  Given the extreme shape of a Kreutz sungrazer's
orbit, its distance from the Sun at aphelion is more than 10$^4$ times (!) greater than at
perihelion, so that perturbations of the angular elements are by far the largest when the
fragmentation event takes place in the general proximity of aphelion, whereas perturbations
of the orbital period skyrocket when splitting occurs at, or close to, perihelion.
Conversely, it is impossible to appreciably perturb the angular elements of a fragment
when the comet breaks up near perihelion, while the orbital period cannot change much on
account of a near-aphelion breakup.  Another peculiar property of Kreutz sungrazers is
their anomalously low orbital velocity at aphelion, of about 20~m~s$^{-1}$, very unusual
at less than 200~AU from the Sun.  As a result, a separation velocity of a few meters per
second could cause a fragment's orbital velocity to differ by more than 10~percent from
the parent's, a sizable relative change that implies a major orbit transformation.  And
if fragmentation events should be distributed randomly in time, their occurrence at large
heliocentric distance is strongly preferred:\ with a 90~percent probability beyond 60~AU
from the Sun and with a 99~percent probability beyond 13~AU.

Marsden (1967) called attention to a tendency of the naked-eye Kreutz sungrazers to arrive
at perihelion in clusters, the two most recent ones about 80~years apart.  Even though he
eventually dismissed this peculiarity as ``largely fortuitous,'' the above lines suggest
that {\it clustering\/} is linked to {\it near-aphelion fragmentation\/}.  On the other
hand, the existence of {\it major temporal gaps\/} between clusters is linked to
{\it near-perihelion fragmentation\/}.  By the same token, it is ruled out in principle
that two Kreutz sungrazers could simultaneously be members of the same population and
the same cluster, even though the strict validity of this rule is mitigated by fragmentation
events that occur far from both perihelion and aphelion,\footnote{The pair of C/1880~C1 and
C/1887~B1, members of Population~I and the 19th century cluster, found to have separated from
their common parent at about 50~AU from the Sun after perihelion (as suggested in Section~5.2),
is a case in point.} as well as by ongoing effects of the outgassing-driven nongravitational
acceleration and possibly by the indirect planetary perturbations.  Yet, compliance with
the rule was clearly demonstrated by C/1882~R1 and C/1965~S1, the members of the {\it same
population\/} belonging to {\it different clusters\/}; and by C/1963~R1, C/1965~S1, and
C/1970~K1, the members of the {\it same cluster\/} belonging to {\it different populations\/}.

\section{Contact-Binary Model for the Progenitor, and Its Cascading Fragmentation}
%
As pointed out in Section 1, the rationale for the new conceptual model is predicated in
part upon the proposed status of C/1843~D1 and C/1882~R1 as the largest surviving masses
of the progenitor.  The $\sim$40~year wide gap between their perihelion times is believed
to be the outcome of the conditions at birth and of the orbital evolution.  As Population~I
is associated with C/1843~D1 and Population~II with C/1882~R1, one would expect a sizable
residual mass associated with Population~Ia to have arrived most probably in the 1860s,
about halfway between 1843 and 1882.  With no sungrazer seen in that window of
time,\footnote{While no bright Kreutz sungrazer was detected in the 1860s, a ``sun-star,''
possibly a daytime sungrazing comet, was reported to be seen in close proximity of the
Sun from the Jiangsu province on July 18, 1865 according to a Chinese compendium of
historical records of celestial objects (Strom 2002).  No further information is however
available, so even if this object was a comet, its Kreutz system membership remains
undetermined. --- Independently, a remote possibility exists that C/1945~X1, which
passed perihelion eight decades after the critical time, could be part of the neck's
residual mass, if the parent had split similarly to the 1882/1965 parent; two of the
orbits Marsden (1967) derived for C/1945~X1 had the longitude of the ascending node
near 352$^\circ$ (equinox J2000), conforming fairly well to the value of Population~Ia.
The comet's orbital uncertainty obviously makes this point very speculative.}
the scenario that one confronts appears to be that of a progenitor comet made up of {\it
two mass peaks with a distinct mass deficit in between\/} --- clearly inviting comparison
with the {\small \bf contact binary} that consists of {\small \bf two sizable lobes}
connected by a {\small \bf narrower neck}.

The idea of a contact-binary progenitor is broadly corroborated by gradually accumulating
evidence based on spaceborne close-up imaging of cometary nuclei and Kuiper Belt objects.
This evidence suggests that the contact binary is a fairly common figure among these
objects.  It suffices to mention two prominent examples:\ comet 67P/Churyumov-Gerasimenko
(e.g., Sierks et al.\ 2015) and (486958) Arrokoth (e.g., Stern et al.\ 2019). 
%
%

\begin{figure}[h]
\vspace{-6.1cm}
\hspace{1cm}
\centerline{
\scalebox{0.74}{
\includegraphics{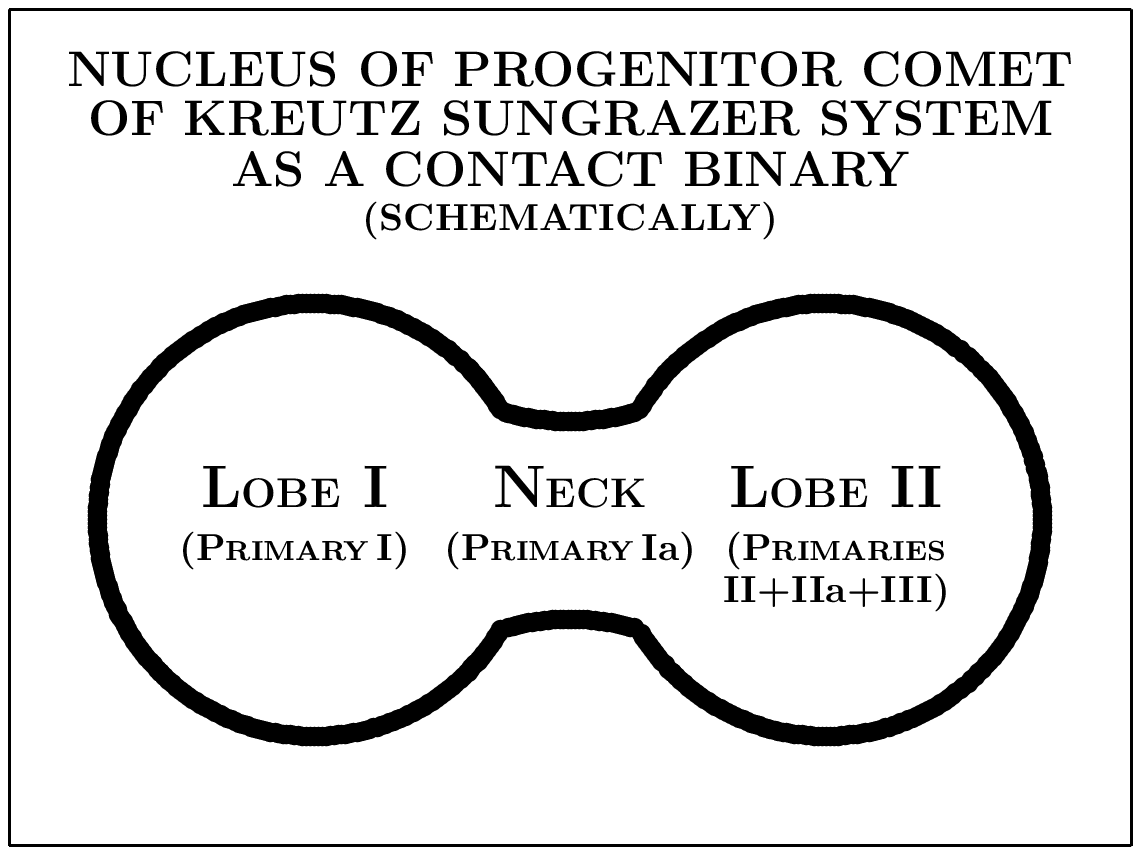}}}
\vspace{-9.6cm}
\caption{Proposed model for the Kreutz system's progenitor sungrazer, shaped as a symmetric
contact binary of uniform density.  The two lobes are assumed to be spherical, of
equal dimensions, and connected with a narrower neck.  Lobe~I consists of Primary~I,
the precursor of Population~I; Lobe~II includes Primaries~II, IIa, and III, the precursors
of Populations~II, IIa, and III, respectively; while the neck --- Primary~Ia --- is the
precursor of Population~Ia.  The unknown precursors of Populations~Pre-I (part of Lobe~I),
IIIa, and IV (parts of Lobe~II) are omitted; Population~Pe, as a side branch of Population~I,
derives from Lobe~I.  The model postulates that in the course of the Kreutz system's initial
fragmentation event the lobes separated from the neck with opposite velocities of equal
magnitude.{\vspace{0.6cm}}}
\end{figure}

To account for all populations of the Kreutz system with naked-eye, massive fragments
of the progenitor $\Re$, I identify Lobe~I with Primary~I, a precursor of C/1843~D1, and
assume that Lobe~II included Primaries~II (a precursor of C/1882~R1 and C/1965~S1), IIa (a
precursor of C/1970~K1), and III (a precursor of C/2011~W3), as shown in Figure~2.  The neck
is identified with Primary~Ia.  The possible existence of Primary~IIa{\large $^{\!\ast}$}
is discussed in Section~5.2.  To account for the populations whose massive fragments are
unknown, one should admit that Lobe~I incorporated the parent to dwarf sungrazers of
Population~Pre-I, whereas Lobe~II the parents to dwarf sungrazers of Populations~IIIa and
IV, which are all omitted from Figure~2.  Population~Pe, as a source of dwarf sungrazers
associated with C/1963~R1, is merely a side branch of Population~I, as already pointed out.

\subsection{Initial Breakup of the Progenitor}  
In the following I assume that the two lobes of the rotating progenitor of uniform density
were spheres of equal size and that the progenitor's center of mass coincided with the
center of mass of the neck, whose figure was symmetric relative to the two lobes (Figure~2).
The progenitor is proposed to have broken up near aphelion, assumed nominally at $\sim$170~AU
from the Sun, into the three fragments, as either lobe separated from the neck at the same
rate in exactly opposite directions, while the neck itself continued to move in the orbit of
the pre-breakup progenitor.  In practice it is of course possible (if not likely) that the
contact binary split into only two fragments, with the neck remaining attached to one of the
lobes (to break off later) or the plane of fissure running through the neck.  It is obviously
easier to accept splitting of a contact binary rather than a quasi-spherical nucleus (as was
the case with the two-superfragment model; Sekanina \& Chodas 2004).
%
%
\begin{table*}
\vspace{-4.2cm}
\hspace{1.55cm}
\centerline{
\scalebox{1}{
\includegraphics{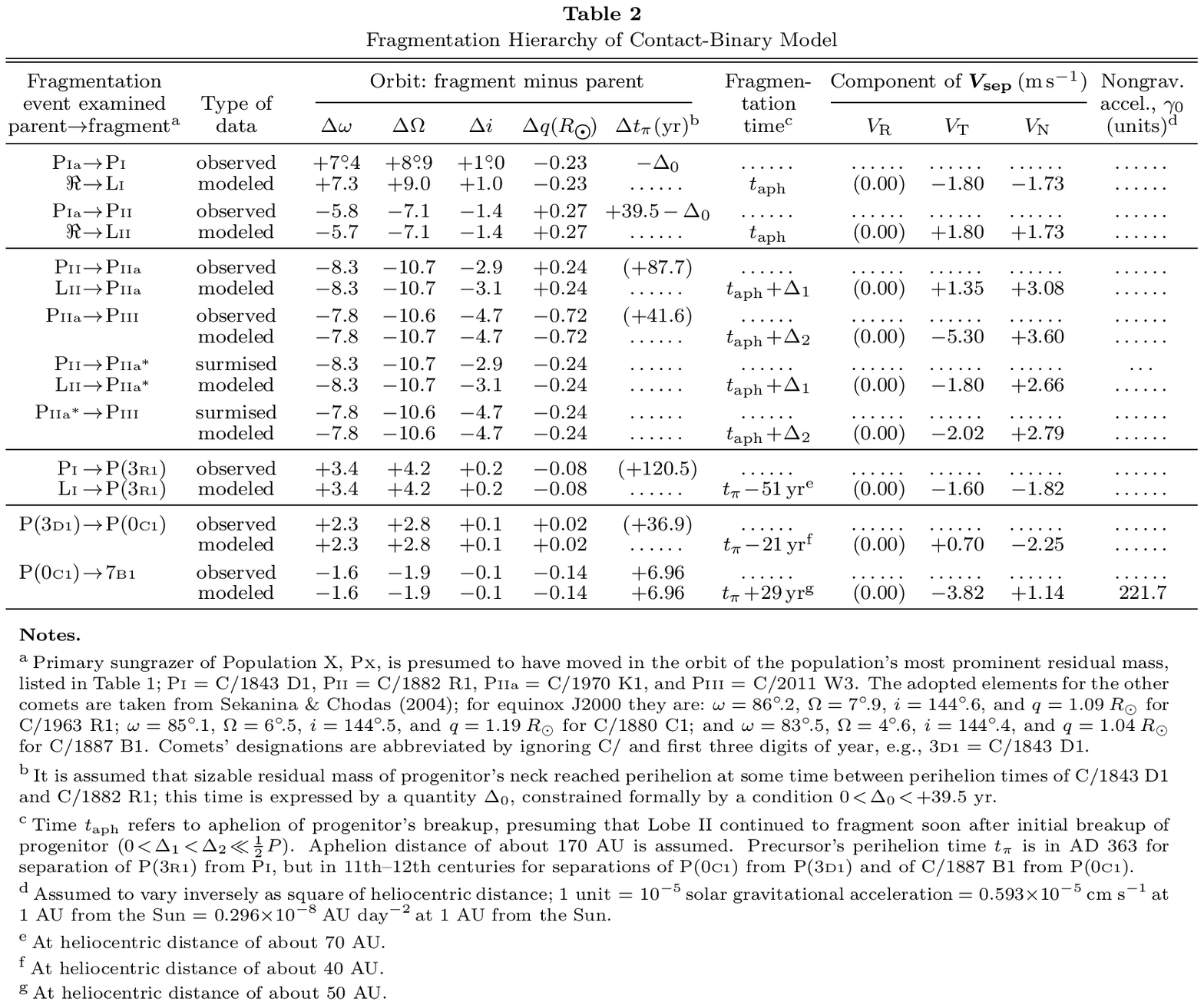}}} 
\vspace{-10.9cm}
\end{table*}

Although nominally unknown, the orbit of the pre-breakup progenitor --- maintained in the
proposed scenario as the orbit of Primary~Ia --- can be constructed from the orbits of Lobe~I
and Lobe~II approximated by the orbits of C/1843~D1 and C/1882~R1, respectively.  In an
effort to find a set of orbital elements for the progenitor $\Re$, I have searched for
(i)~a particular value of the out-of-orbit (normal) component of the separation velocity
such that it fits the differences in $\omega$, $\Omega$, and $i$ between Primaries~Ia and
I, and that, at the same time, this velocity component of equal magnitude but opposite
sign fits the differences in the three elements between Primaries~Ia and II; and (ii)~a
particular value of the transverse component of the separation velocity such that it fits
the difference in $q$ between Primaries~Ia and I, and that, at the same time, this velocity
component of equal magnitude but opposite sign fits the difference 
in $q$ between Primaries~Ia and II.\footnote{The separation velocity is customarily
expressed in an RTN coordinate system referred to the parent's center and its orbital
plane; the N axis points to the north orbital pole, the R axis away from the Sun; and the
T axis in a direction such that RTN is a right-handed orthogonal system.  The differences
in the orbital elements triggered by the separation velocity are found by iterating the
following procedure:\ (1)~convert the orbital elements of the parent to the ecliptic
coordinates of its radius vector and orbital-velocity vector at the time of breakup;
(2)~derive the ecliptic components of the orbital-velocity vector of the fragment by
converting the components of the separation velocity vector from the RTN system to the
ecliptic system and adding them to the parent's orbital-velocity vector; (3)~convert
the ecliptic components of the fragment's radius vector (equal to the parent's radius
vector) and its orbital-velocity vector to its orbital elements; and (4)~subtract the
fragment's elements from the parent's elements to derive the perturbations as effects
of the separation velocity vector.  If the derived orbital perturbations do not fit
the differences in the elements, iterate steps~(2) to (4) with improved values for the
separation velocity vector components until convergence has been reached.}  The radial
component of the separation velocity vector could not be determined because it primarily
contributed to the perihelion arrival time, which, as a matter of course, is also subject
to the indirect planetary perturbations and nongravitational forces.

The first four rows of Table 2 refer to the initial breakup of the progenitor $\Re$.  The 
large perturbations of Lobes~I and II in the angular elements, of up to nearly 10$^\circ$, are
fitted with an out-of-plane component of the separation velocity of less than 2~m~s$^{-1}$
and, similarly, the fairly sizable perturbations of the lobes in the perihelion distance,
of nearly 0.3~{\Rsun}\,, are fitted with a transverse component of the separation velocity
of less than 2~m~s$^{-1}$.  Thus, very modest rates of separation readily explain nominally
large orbital differences between Populations~I and II, in line with the general arguments
in Section~4.  The solution for the angular elements is particularly impressive, because
a {\it single\/} parameter closely approximates the effects in three elements of either
lobe, i.e., six variables!  The tabulated data also show that the symmetry of the lobes'
separation velocity does not imply a symmetry of the resulting perturbations of the elements.

I now illustrate this exercise on the longitude of the ascending node.  Table~1 shows that
\mbox{$\Omega = 3^\circ\!$.7} for Primary~I in the column {\it main\/} of Population~I and
\mbox{$\Omega = 347^\circ\!$.7} for Primary~II in the column {\it main\/} for Population~II,
so that the difference Primary~II minus Primary~I equals $-16^\circ\!$.0.  For the progenitor's
breakup at aphelion, the best orbital model for Primary~Ia, found by trial and error,
included Lobe~I's out-of-plane separation velocity of $-$1.73~m~s$^{-1}$ (and Lobe~II's
out-of-plane separation velocity of +1.73~m~s$^{-1}$).  This model's longitude of
the ascending node for Primary~Ia was 354$^\circ\!$.8, so that Primary~I's
value of $\Omega$ differed from it by +8$^\circ\!$.9, while Primary~II's value of $\Omega$ by
$-7^\circ\!$.1, as shown, respectively, in the first and third rows of Table~2.  The model
predicted that, relative to the progenitor $\Re$, the longitude of the ascending node of
Lobe~I should have been perturbed by +9$^\circ\!$.0 and of Lobe~II by $-7^\circ\!$.1, as
shown, respectively, in the second and fourth rows of Table~2.  The model's fit to $\Omega$ is
thus perfect for Lobe~II and good to 0$^\circ\!$.1 for Lobe~I.

\subsection{Subsequent Fragmentation Events}  
Not counting minor pieces, the initial breakup is expected to have nominally resulted in two
or three fragments:\ the two lobes and, possibly, the separate neck, with all or nearly all
mass concentrated in the lobes in either case.  The event is likely to have been chaotic and
its products certainly not perfectly symmetric, which could in part explain the average orbit
of the dwarf sungrazers in Population~Ia deviating slightly from the orbit of the progenitor's
symmetric model (Table~1).

In line with the proposed classification of the Kreutz system (Section~2), it is contemplated
that it was Lobe~II that, following the initial breakup, continued to fragment far from the
Sun probably during the same orbit.  In the earliest secondary fragmentation event, most of
the mass of Lobe~II became Primary~II (a precursor of C/1882~R1 and C/1965~S1), a smaller part
in a detached fragment consisted of the future Primaries~IIa (a precursor of C/1970~K1), III (a
precursor of C/2011~W3), and hypothesized primaries of Populations~IIIa and IV.  The transverse
component of the separation velocity of Primary~IIa from Primary~II was again less than
2~m~s$^{-1}$, the out-of-orbit component almost exactly 3~m~s$^{-1}$, both deemed plausible.
Lobe~I appears to have been more susceptible to fragmentation at smaller heliocentric distances.

Given the large difference between the perihelion distances of C/1970~K1 and C/2011~W3, there
are two possibilities for explaining the parent to C/2011~W3.  If a separation velocity
exceeding 5~m~s$^{-1}$ is not considered unacceptably high, Primary~III may have separated
from Primary~IIa, perhaps soon after the breakup of Lobe~II.  If this velocity is deemed
unacceptably high, one can contemplate that, in addition to Primary~IIa, another fragment
that I~refer to as Primary~IIa{\large $^\ast$} was released from Lobe~II into an orbit
with a perihelion distance {\it smaller\/} than that of Lobe~II, and that Primary~III was
subsequently released from Primary~IIa{\large $^\ast$}.  In this scenario the separation
velocity would remain within about 3~m~s$^{-1}$, comparable to those from the progenitor and
Lobe~II.  Possible fragmentation scenarios involving Lobe~II and Primaries~IIa, IIa{\large
$^\ast$}, and III are listed in rows \mbox{5--12} of Table~2.
%
%
\begin{figure}[t]
\vspace{-1cm}
\hspace{2.75cm}
\centerline{
\scalebox{0.77}{
\includegraphics{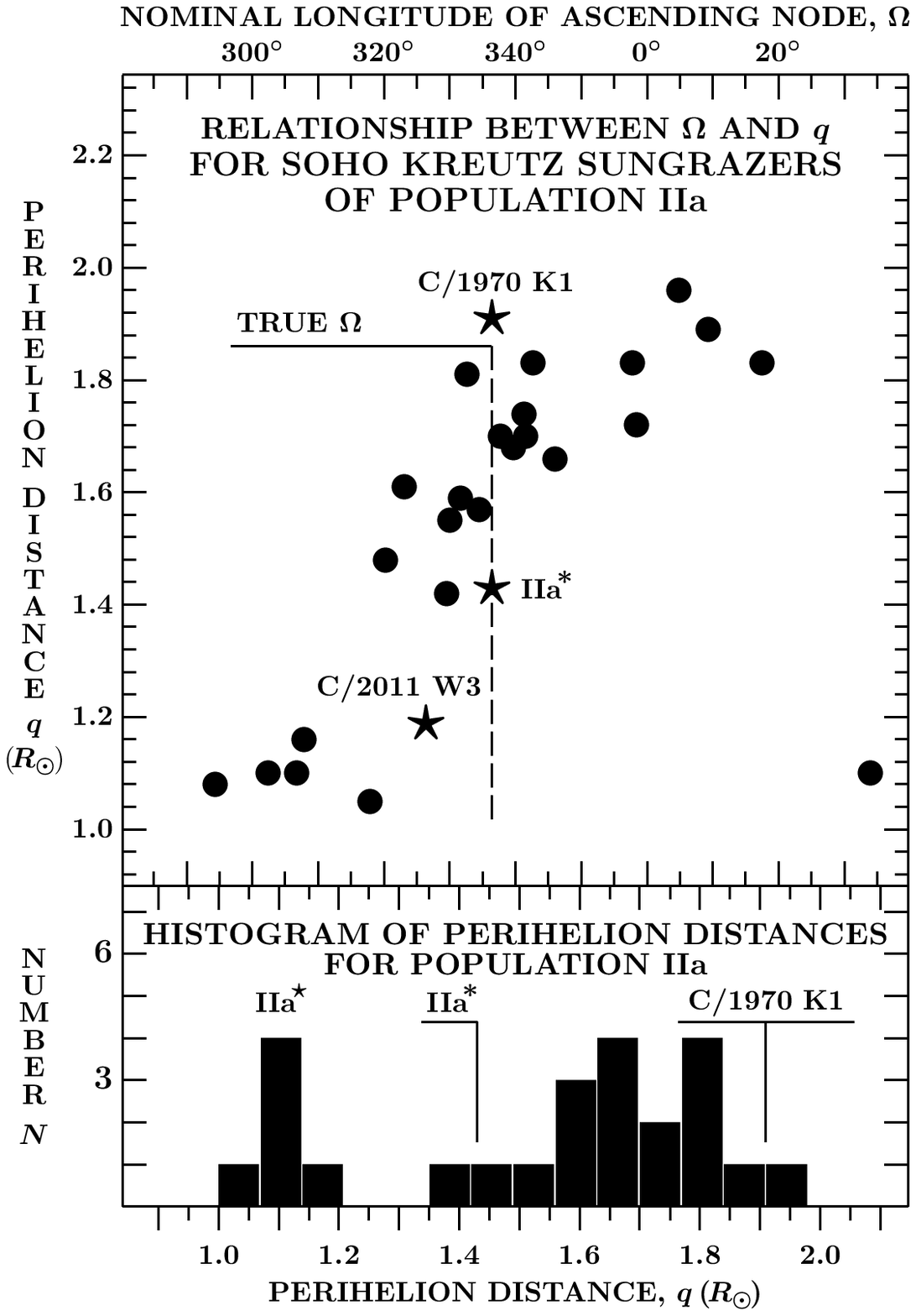}}}
\vspace{-9.5cm}
\caption{{\it Upper panel\/}:\ Relation between the nominal longitude of the ascending
node, $\Omega$ (equinox J2000), and the perihelion distance, $q$, for 24~dwarf Kreutz
sungrazers of Population~IIa imaged exclusively by the C2 coronagraph of the SOHO
Observatory.  The data are sharply split into two subsets:\ (i)~a major one with
\mbox{$q > 1.4\:R_\odot$}, which{\vspace{-0.08cm}} includes objects in orbits ranging
from those similar to C/1970~K1 to those similar to Primary~IIa{\large $^\ast$} (see
the text); and (ii)~a minor one with{\vspace{-0.08cm}} \mbox{$q < 1.2\:R_\odot$}, which
contains objects in orbits of a type referred to in the text as Primary~IIa{\large
$^\star$}; they may be part of transition to Population~III (C/2011~W3).  {\it Lower
panel\/}:\ Histogram of perihelion distances of the 24~dwarf{\vspace{-0.06cm}} sungrazers,
showing a sharp peak of Primary~IIa{\large $^\star$} and a broad bulge of the major
subset, whose boundaries are{\vspace{-0.06cm}} approximately delineated by the
perihelion distances of C/1970~K1 and Primary~IIa{\large $^{\!\ast}$}.{\vspace{0.4cm}}}
\end{figure}

Primary~IIa{\large $^\ast$} may not be a purely theoretical construct.  To examine its
potential footprint in the data, I plot, in the upper part of Figure~3, the perihelion
distance against the nominal longitude of the ascending node for the 24~dwarf Kreutz
sungrazers of Population~IIa.  The distribution of objects in the plot is startling:\
18~points cluster in an area between 320$^\circ$ and 20$^\circ$ in the nodal longitude and
between 1.4~{\Rsun} and 2.0~{\Rsun} in the perihelion distance, while the six remaining
points are located completely outside the area of the cluster, confined to a narrow interval
of perihelion distances between 1.0 and 1.2~{\Rsun}\,, thus closely matching the perihelion
distance of C/2011~W3.  Since the nominal nodal longitudes are merely a measure of the
nongravitational effect and all refer to nearly the same true nodal longitude of about
336$^\circ$ (Table~1), the plot at the top of Figure~3 effectively collapses into a histogram
displayed in the lower panel.  The cluster now becomes a very broad bulge on the right, while
the six isolated points make up a sharp peak on the left; the two features are separated by a
wide gap with no sungrazers between 1.2 and 1.4~{\Rsun}\,.  Comet{\vspace{-0.06cm}} C/1970~K1
and Primary~IIa{\large $^\ast$}, introduced above, pinpoint approximately the boundaries
of the bulge, while the sharp peak can be equated with debris of what I refer to as
Primary~IIa{\large $^\star$}.  Figure~1 shows that these six objects with anomalously small
perihelion distance fit the same relationship between $\Omega$ and $B_\pi$ as the other
members of Population~IIa, but their nominal nodal longitudes are more distant from the
true value of $\sim$336$^\circ$.  It is not clear whether comet C/2011~W3 and Primary~III
are more closely linked to Primary~IIa{\large $^\ast$}, as proposed in Table~2, or to the
hypothetical Primary~IIa{\large $^\star$}.  This case serves to illustrate the enormous
complexity of the Kreutz system's structure.

On the other hand, Table 2 offers a fairly straightforward model for the history of comet C/1963~R1,
which was a major hurdle to Marsden's (1989) fragmentation scenario.  Rows \mbox{13--14}
suggest that the comet's precursor, P(3{\scriptsize R1}), separated, after the progenitor's
breakup, from Lobe~I on its path to perihelion, which it should reach almost simultaneously
with the other fragments of the progenitor (Section~5.3).  Since the whereabouts of
P(3{\scriptsize R1}) depend on the history of the Kreutz system as a whole, a description
of its further evolution is postponed to Section~5.3.  
 
Although not tabulated, Primary~III is the likely parent to Primary~IIIa and Primary~IIIa
the likely parent to Primary~IV.  Preliminary computations suggest that the out-of-orbit
separation velocities would amount to 3.7~m~s$^{-1}$ and 3.4~m~s$^{-1}$, respectively.  The
remaining two rows of Table~2 show that the applied technique fits equally well more recent
fragmentation events nearer perihelion, in which the parent to C/1880~C1, P(0{\scriptsize
C1}), split off from the parent to C/1843~D1, P(3{\scriptsize D1}), and, later still,
C/1887~B1 split off from the parent it had in common with C/1880~C1.  It is noted that the
timing of the tabulated fragmentation event P(3{\scriptsize D1})$\rightarrow$P(0{\scriptsize
C1}) implies that X/1106~C1 should have arrived at perihelion double.  However, as long as
the radial component of the separation velocity was much smaller than 1~m~s$^{-1}$, the main
comet and the companion passed perihelion only a fraction of an hour apart.

\subsection{Pedigree Chart, Key Historical Sungrazers, and Estimate of the Kreutz System's
 Age} 
The products of a sequence of fragmentation events are displayed in a {\small \bf
pedigree chart}, which describes possible relationships among fragments of successive
generations.  Because the perceived relationships are not necessarily unambiguous, neither
is the pedigree chart.  The chart in Figure~4 excludes Populations~Pre-I, IIIa, and IV
because no naked-eye sungrazer is known to be associated with them.\footnote{Population~Ia
is included because of its special status.}  For the same reason the chart does not single
out historical precursors of the 19th--21st century naked-eye members of the Kreutz system
that the classification scheme is predicated upon.

Of historical objects, the widely recognized candidate for the Kreutz system membership is
the extensively observed comet X/1106~C1.  Even though no orbital elements could be derived,
its reported motion across the sky was deemed consistent with the expected path of a Kreutz
sungrazer by Hasegawa \& Nakano (2001).  However, whether X/1106~C1 was a parent to C/1882~R1
and C/1965~S1 (i.e., Population~II) or to C/1843~D1 and other members of Population~I remains
controversial.  The former possibility was contemplated by Kreutz (1888, 1901), advocated by
Marsden (1967), and independently considered by Sekanina \& Chodas (2004); in Figure~4 the
comet would be identical with Precursor~C.  Marsden predicated this preference on his result
of integration back in time of the accurately-determined orbit of C/1965~S1.  He derived
September 1116 as{\vspace{-0.04cm}} the time of the comet's previous perihelion\footnote{This
time was obtained by integrating the nonrelativistic orbit of nucleus~A of comet C/1965~S1; a
slightly different time resulted by integrating the relativistic orbit.} and noted that
this was ``remarkably close to February 1106,'' the arrival time of X/1106~C1.  At first
sight, the argument looks reasonable.  But what if Marsden's orbital computations were
more accurate than he himself believed?  This is a distinct possibility given the very
small error of the comet's orbital period, amounting to just $\pm$2.1~years and placing
X/1106~C1 at 5$\sigma$.

Comet C/1882 R1 was less helpful because at perihelion it split into six major fragments
(Kreutz 1888) and its pre-split orbit was quite uncertain.  Marsden's (1967) integration
of Kreutz's (1891) nonrelativistic orbit for nucleus B gave the previous perihelion time
in April 1138, and the eccentricity had to be adjusted to make the time coincide with comet
C/1965~S1 in order to determine the differences between the orbital elements of the two
comets in the early 12th century.  It turned out that the differences closely mimicked
those between the two nuclei of C/1965~S1, a coincidence that Marsden regarded as virtual
proof that C/1882~R1 and C/1965~S1 split off from their shared parent in the early 1100s,
regardless of whether or not it was X/1106~C1.  Unfortunately, this argument is not valid
because the orbital differences between the two fragments of C/1965~S1 were products of
uncertainties in the positional data used to compute the fragments' orbits, not of the
conditions at the time of fragmentation.  As the differences amounted to about 0$^\circ\!$.01
in the angular elements and about 0.001~{\Rsun} in the perihelion distance, the separation
velocity required to fit them if the breakup occurred within hours of perihelion equals
50--100~m~s$^{-1}$ or more and is thus hopelessly much too high.  The bottom line is that
the two comets did split off from a common parent at some time in the early 12th century,
but not because of the correlation between the differences in the orbital elements.

%
\begin{figure*}[t]
\vspace{-4cm}
\hspace{-0.22cm}
\centerline{
\scalebox{0.877}{
\includegraphics{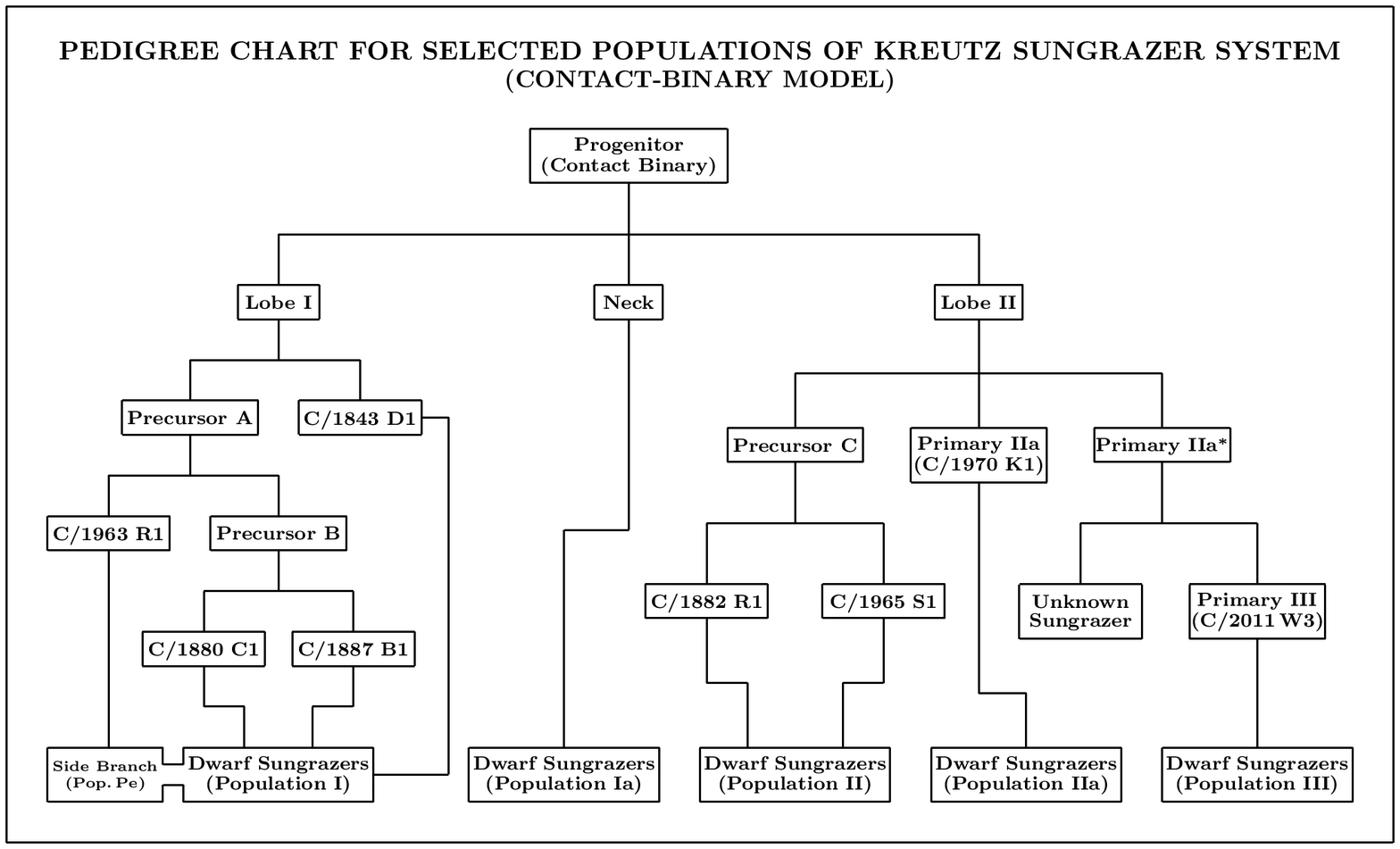}}}
\vspace{-10.9cm}
\caption{Example of a pedigree chart for the contact-binary progenitor assumed to
have broken up into three pieces:\ two lobes and the neck.  Based on evidence from
the naked-eye Kreutz system members in the 19th--21st centuries, the scenario shows
that Lobe~I split into C/1843~D1 and a fragment (referred to as Precursor~A) that
subsequently split again into C/1963~R1 and another fragment, Precursor~B or
the parent to C/1880~C1 and C/1887~B1.  The debris of comet C/1843~D1 is detected as
SOHO dwarf comets of Population~I, whereas the debris of C/1963~R1 ended up in
Population~Pe, a side branch{\vspace{-0.01cm}} to Population~I.  The scenario for Lobe~II
includes three major fragments:\ one was a parent to C/1882~R1 and C/1965~S1, referred
to as Precursor~C; the{\vspace{-0.1cm}} others were Primaries~IIa, surviving as
C/1970~K1, and a hypothetical Primary~IIa{\large $^\ast\!$} that is presumed to have
later broken up into Primary~III, surviving as C/2011~W3, and an unknown sungrazer.
The debris of Precursor~C includes SOHO dwarf sungrazers in Population~II, while the
debris{\vspace{-0.01cm}} of Primaries~IIa and III contains dwarf sungrazers in
Populations~IIa and III, respectively.{\vspace{0.6cm}}}
\end{figure*}

On the other hand, Marsden made a point by arguing that because equating the post-split
position of the center of mass of C/1965~S1 with the primary nucleus was an approximation,
the orbit integration back over more than eight centuries could involve additional hidden
uncertainties, not reflected in the formal mean error of the orbital period.  This line
of reasoning was further pursued by Sekanina \& Chodas (2002) in their effort to more
closely examine whether X/1106~C1 was indeed the shared parent of C/1882~R1 and C/1965~S1.
They adjusted the orbital position of the center of mass of either comet relative to the
positions of the observed fragments to match the presumed perihelion time of X/1106~C1,
1106~January~25 according to Hasegawa \& Nakano (2001), and interpolated the orbital
elements proportionately.  If X/1106~C1 was the parent, the positions of C/1882~R1 and
C/1965~S1 should have essentially merged at the point of fragmentation.  One of course
expects that the merging should have occurred at, or very close to, perihelion.  Instead,
Sekanina \& Chodas found that the positions of the two comets were nearest each other
18~days after perihelion, at 0.75~AU from the Sun, when the orbital velocities of the nascent
fragments differed by 7~m~s$^{-1}$.  Since these results are incompatible with the events
of fragmentation observed in both 1882 and 1965, one could interpret the anomalous solution
for the 1106 breakup as a product of incorrect starting assumptions, i.e., by admitting that
X/1106~C1 was {\it not\/} the parent to C/1882~R1 and C/1965~S1.

The alternative scenario, in which X/1106~C1 was assumed to be the parent to, or the preceding
appearance of, C/1843~D1, was adopted in Sekanina \& Chodas (2007).  In this case, the parent
to C/1882~R1 and C/1965~S1 should have reached perihelion years after X/1106~C1.  Either way,
two bright Kreutz sungrazers should have arrived within two decades or so of each other in
the late 11th to early 12th century, but only one was recorded.  Whether the other arrived
before or after X/1106~C1, the chance is that it passed perihelion in broad daylight
between late May and mid-August and was missed.\footnote{Because of the orientation of its
orbit in space, a Kreutz sungrazer approaches and recedes essentially from behind the Sun
between late May and mid-August.  When passing perihelion within weeks of late December,
a Kreutz sungrazer is confined to high southern declinations and stays below the horizon
for most north-hemisphere observers, except when near the Sun in broad daylight.}  Another
dilemma, the whereabouts of the surviving mass of the progenitor's neck, is of lesser concern
because of the possibility of its prior disintegration or sticking, at least temporarily,
with one of the progenitor's two lobes, as already pointed out.

To pinpoint the parents to X/1106~C1 and its missing sibling, which should have arrived
at perihelion in the 4th or 5th century, is extremely difficult.  Sekanina \& Chodas
(2007) suggested that the comets of AD~423 or AD~467 might have been good candidates,
as both objects were ranked fairly high as potential Kreutz sungrazers by England (2002)
and the first was also on Hasegawa \& Nakano's (2001) list of Kreutz suspects.  However,
a linkage of either of the two comets with X/1106~C1 showed that the subsequent return to
perihelion occurred much too early and that a high nongravitational acceleration would be
required to significantly stretch the orbital period in order to fit the perihelion time
of C/1843~D1.  Another problem with the two 5th century candidates is of the same nature
as with the 11th/12th century sungrazers:\ for neither the comet of 423 nor 467 is there
any record of a second sungrazer (or two more sungrazers to account for the neck), passing
perihelion at most a~year or so --- but possibly as little as days --- apart.  Interestingly,
in the Sekanina \& Chodas (2004) model the two superfragments passed perihelion one week
apart in AD~356 and six years apart in the beginning of the 12th century.  However, the
two-superfragment model allowed separation velocities of up to $\sim$10~m~s$^{-1}$, much
higher than contemplated in this paper.

Given the disappointing experience with the 5th century contenders, I now inspected the
general catalogue of ancient and medieval comets by Kronk (1999), which is Kreutz system
indifferent.  I focused my attention on the critical period of time centered crudely on
the middle of the 4th century, between AD~340 and 370,\footnote{This choice was partly
affected by the results of the paper on the two superfragments
(Sekanina \& Chodas 2004), which indicated their first perihelion passages one week
apart in AD~356.} and found two possible candidates, presented below.  One is much more
promising than the other, but very little information is available on either.

One candidate is the comet of AD~349, extending into 350.  Listed, next to Kronk, by both
Ho (1962) and England (2002), this could be a case of three independent objects separated
by two-month wide gaps.  If so, only the comets at either end might be Kreutz members,
but not the middle one.  Of interest is Kronk's comment that the preceding comet in his
catalogue, in the year 343, was almost exactly six years earlier observed at the same
location as the first of the three sightings in 349/350.  On the other hand, England
assigned the 349 entry low Kreutz-system membership ranking and did not list the 343
comet at all.  Hasegawa \& Nakano (2001) did not include either one in their table of
Kreutz suspects, probably because of the lack of adequate data that they needed to test
compatibility with the sungrazing orbit.  Yet, the pair of potential sungrazers in 349/350
offers a possible scenario for the first appearance of the progenitor's separated lobes,
with the whereabouts of the neck still uncertain.  The four-month wide gap would imply
a radial component of the separation velocity of about 1~m~s$^{-1}$ in the opposite
directions for the two lobes, if the progenitor fragmented close to its previous passage
through aphelion.

It is possible that the comet of 343 was an early fragment of the progenitor.  For
example, a fragment separating from the comet of 372~BC at about 3~AU from the Sun after
perihelion (within $\sim$5~months or so) with a radial component of the separation velocity
of $-$1~m~s$^{-1}$ should indeed have arrived at perihelion about 6~years prior to the two
lobes.  In the following returns to perihelion the fragment would continue to masquerade
as a piece of the neck, because for a breakup at 3~AU the transverse and normal components
of the separation velocity of a few meters per second would cause only very minor changes
in the perihelion distance and the angular elements relative to the progenitor's.

The other candidate that could deliver X/1106 C1 and its missing sibling from the early 12th
century arrived in AD~363.  It was not the eastward traveling comet seen in China between
late August and late September (Ho 1962), at the time of the year when a Kreutz sungrazer
at its brightest would be moving in the southwesterly direction.\footnote{Hasegawa (1979)
suggested that this was the previous appearance of C/1969~Y1 (Bennett), but Seargent (2009)
argued that the comet would have been too faint to detect with the naked eye.}  Rather, what
caught my attention was the {\it plural\/} in an intriguing (though brief and vague) remark by
Ammianus Marcellinus, a Roman historian, who put on record that {\it ``in broad daylight comets
were seen''\/} --- not{\it a\/} comet.\footnote{Barrett (1978) erroneously dated another brief
note by Ammianus on ``comets blazing'' to AD~364.  Coming from the historian's Book~XXX (Rolfe
1939), which begins with the year 374, the reference may pertain to (in part?) the comet of
375, as pointed out by Ramsey (2007).}  All four secondary sources that I have consulted ---
Barrett (1978), Kronk (1999), Ramsey (2007), and Seargent (2009) --- offer the quote, but only
Ramsey and Seargent comment on the time of sighting.  Unfortunately, they differ:\ Ramsey
says August to September~363, whereas Seargent claims {\it late that year\/}, observing that
Kreutz sungrazers ``very late in the year would have had a strong southerly declination and
might have been seen from Italy only in the daytime close to perihelion.''  As Ramsey mentions
``possible corroborating evidence from Asia,'' it is obvious that the time slot he provides
comes from the Chinese comet dismissed above.  Inspection of Ammianus' Book~XXV, translated
by Rolfe (1940), indicates that the narrative about the daylight comets {\it follows\/} his
description of the arrival of Roman Emperor Jovian and his army at Antioch on the Orontes in
the aftermath of the Battle of Samarra in the Sasanian Empire and the sudden death of Emperor
Julian.  It is known that Jovian, accompanied by Ammianus, entered Edessa in September (Elton
2018) and Antioch in October 363 (Lenski 2002).  Ammianus stayed at Antioch, while Jovian
left the town (``in the dead of winter'' according to the historian) to continue his ill-fated
journey to Constantinople.  Thus, August or September could not be the time of sighting of the
daytime comets.  Rather, the celestial splendor was witnessed by Ammianus at Antioch either
toward the end of October or later still in 363.  The location was in fact unimportant, as
Seargent's argument applies to much of the northern hemisphere, not just Italy.  One can guess
that the event was a grand daytime spectacle, whose path across the sky could have resembled
that of Lovejoy's sungrazer in 2011.\footnote{While accepting the comets' steep southerly
%
plunge after peri\-helion, one feels a little uneasy about the absence of records of {\it
independent\/} daytime detections (the case of the comet in 302 shows that Chinese did document
such events in those times; Ho 1962, England 2002, Seargent 2009) that would corroborate
Ammianus' claim.  Yet, his elaborate narrative on comets, which accompanies the remark on the
sighting (Rolfe 1940, Kronk 1999) and touches upon the ideas of Aristotle, Pliny the Elder,
and Plutarch among others, suggests that Ammianus was knowledgeable enough in the subject
and that it would be unwise to question the veracity of his startling statement.  Equally
reassuring was the eloquent description of a ``falling star'' elsewhere in his writings.
Modern historians praise Ammianus' works as factually accurate and admire his impartiality
of judgment and breadth of view.  On the other hand, he did believe in omens and portents
and was a protagonist of paganism in the time when Christianity was elevated to the state
religion of the Roman Empire.}
%
%
\begin{table*}[t]
\vspace{-4.1cm}
\hspace{0.55cm}
\centerline{
\scalebox{1}{
\includegraphics{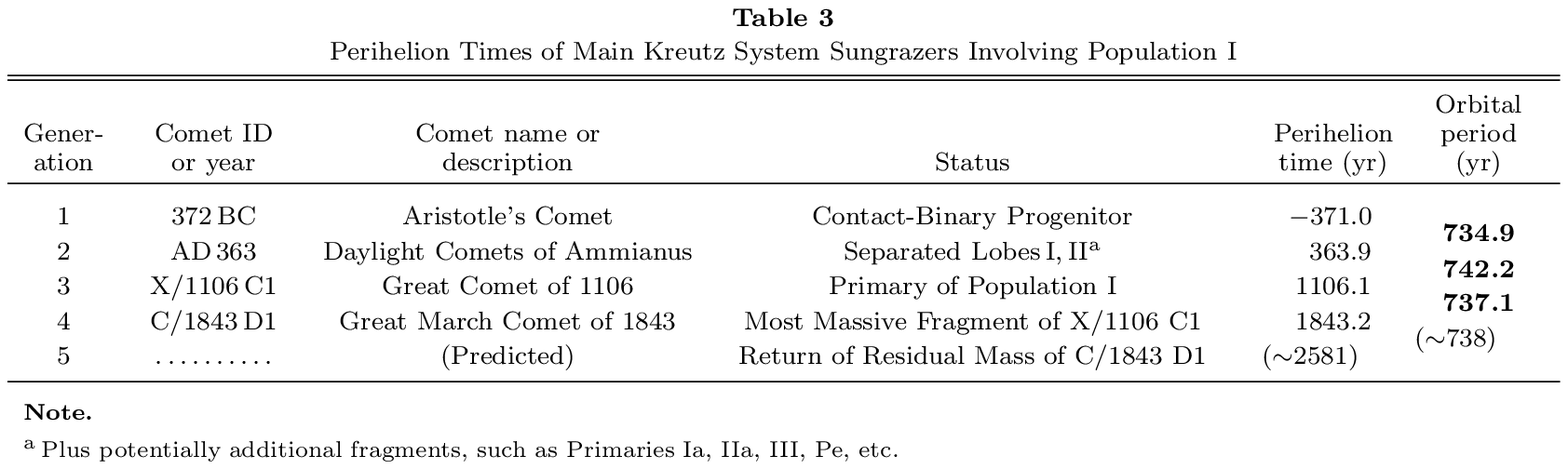}}}
\vspace{-20cm}
\end{table*}

Remarkably, Ammianus' daylight comets satisfy three exacting conditions at once:\ (i)~the
multitude of fragments, whether or not appearing essentially simultaneously, readily avoids
the disconcerting dilemma of missing sibling(s); (ii)~the daytime sighting is consistent with
the comets' expected exceptionally great brightness (X/1106~C1, C/1843~D1, and C/1882~R1 were
all seen in broad daylight near perihelion); and (iii)~the year of arrival offers a point in
time that fits just about perfectly the sequence of perihelion returns of the main surviving
mass of Lobe~I:\ from the progenitor in the year 372~BC;\footnote{This comet is believed to
have appeared in the winter, as 373~BC turned into 372~BC; assuming a perihelion time in late
December or early January, \mbox{$t_\pi = -371.0$}.} to an essentially concurrent appearance
with Lobe~II in AD~363;\footnote{For late in the year, I adopt a perihelion time of
\mbox{$t_\pi = 363.9$}.} to the first separate show as X/1106~C1;\footnote{According
to Hasegawa \& Nakano (2001), the perihelion time was \mbox{$t_\pi = 1106$ Jan $25 \pm 5 =
1106.1$}.} and, finally, to the most recent, equally impressive display as C/1843~D1 (for which
\mbox{$t_\pi=1843.2$}); it is predicted to return late in the 26th century.  As seen from
Table~3, the orbital periods between the consecutive generations are nearly constant and the
%
%
variations so small that they probably could be accounted for by the indirect planetary
perturbations.  The comets of 363 appear to be a distinctly better choice for the Kreutz
sungrazers than the comets of 349, anchoring X/1106~C1 firmly in Lobe~I (i.e., Population~I).
I might venture to suggest that {\it if it were not for Ammianus' utter failure to attend
to any detail whatsoever in recording this celestial episode, his daylight comets could
have become the smoking gun\/} that I have long been looking for.

The motion of the main residual mass of Lobe~II --- a precursor of C/1882~R1, C/1965~S1,
C/1970~K1, and C/2011~W3 --- is in this scenario related to the motion of Lobe~I by assuming
a differential radial nongravitational acceleration driven by the sublimation of water ice,
whose magnitude is given by Marsden et al.'s (1973) parameter of \mbox{$A_1 = +0.93 \times
\!10^{-10}$\,AU day$^{-2}$}.  It implies that Lobe~II was about 50~km across.  The missing
sibling of X/1106~C1 would have passed perihelion most probably in mid-1119, implying for
the two most recent orbital periods of the residual mass of Lobe~II about 756~yr and 763~yr,
respectively.  It would be predicted to return in the second half of the 27th century, about
80~years after the main residual mass of Lobe~I, but because C/1882~R1 was subjected to
extensive fragmentation, the debris will be returning over several centuries late in the
third millennium.  

The proposed model is apt to fit a wide range of possible fragmentation scenarios for the
contact-binary progenitor.  While Ammianus noted neither the number of comets seen in broad
daylight nor the extent of their appearance in time, the model is consistent with up to
at least seven major fragments (Primaries~I, Ia, II, IIa, III, IIIa, and IV), not counting
additional, smaller ones with shorter lifetimes.  In AD~363 the lobes should have arrived only
days apart, if the radial component of their separation velocity was{\vspace{-0.04cm}} very low
(much lower than 1~m~s$^{-1}$), but months{\vspace{-0.04cm}} apart for a separation velocity of
about 1~m~s$^{-1}$ for a breakup at aphelion.  The time gap between the arrivals of the lobes
should be greater, if the progenitor broke up before aphelion, but smaller if after aphelion.
The differences in the angular orbital elements constrain the time of fragmentation event to
within about 300~years of aphelion, at more than $\sim$100~AU from the Sun, if the separation
velocity was to be kept well below 5~m~s$^{-1}$.  A nearly simultaneous arrival at the subsequent
perihelion should imply the separation velocity vector essentially in the plane normal to
the radius vector and the spin axis pointing approximately at the Sun.

In terms of the propensity to {\it bulk\/} fragmentation, Lobe~I and Lobe~II were not alike.
The disparity is apparent in Figure~1, in which the longitudes of the ascending node for
Populations~I, Pe, Pre-I, and (with the usual caveat) Ia, related to Lobe~I, cluster closer
to one another; whereas for Populations~II, IIa, III, IIIa, and IV, linked to Lobe~II, the
nodal longitudes are stretched wider apart.  Quantitatively this unevenness is reflected in
the range of nodal longitudes:\ 14$^\circ$ for Lobe~I and 44$^\circ$ for Lobe~II (Table~1).
Because of the dependence of the nodal-longitude perturbations on heliocentric distance
(Section~4), the discrepancy is easy to interpret:\ the material making up Lobe~I fragments
closer to the Sun than the material making up Lobe~II.  Yet, Table~2 shows that {\it
closer\/} does not mean {\it close\/}:\ the Population~I comets separating from their
common parent with C/1843~D1 appear to have done so at estimated heliocentric distances
of 40--70~AU, less than one half the distance to aphelion.  On the other hand, C/1882~R1, the
presumed main residual mass of Lobe~II, and C/1965~S1 separated from their common parent
right at perihelion (Marsden 1967), as did the six secondary nuclei of C/1882~R1 and the
nuclear pair of C/1965~S1.  This behavior contrasts with C/1843~D1, which was never observed
double or multiple, even though the dominant contribution of Population~I (associated with
C/1843~D1) to the SOHO dwarf sungrazers could mean that the tendency of the Lobe~I material
was to fragment directly into smaller-size debris.

A major fragment whose birth is in the proposed fragmentation scenario linked to Lobe~I is
C/1963~R1.\footnote{This is contrary to the two-superfragment model of Sekanina \& Chodas
(2004), in which C/1963~R1 derived from Superfragment~II, equivalent here to Lobe~II.}  Its
early history, outlined briefly in Section~5.2, ended with a precursor P(3{\scriptsize R1})
arriving at perihelion essentially simultaneously with Lobe~I.  To match the scenario
predicated on Ammianus' brief account, the precursor should have reached perihelion in late
AD~363, contributing to the show of daylight comets and being a subject to more fragmentation
at that time.  One of the precursor's fragments is predicted to have been thrust into an
orbit with a period of 678~yr, having acquired a minor change in the momentum equivalent to
\mbox{$\Delta P = -57$ yr}.  This object is then envisioned to have returned to perihelion
as one of the two comets of AD~1041 (England 2002), probably the Korean-Byzantine one,
observed first on September~1 and listed as a Kreutz comet suspect by Hasegawa \& Nakano
(2001).  New fragments were given birth by the comet at this return to perihelion.  A
significant change in the momentum of one fragment, equivalent to \mbox{$\Delta P = +244$
yr}, would have moved it into an orbit with a period of 922~yr, making it to return to
perihelion as C/1963~R1.\footnote{The reader who deems this change in the orbital period
excessively large is reminded that the results of Kreutz's (1891) orbital computations
implied \mbox{$\Delta P = 204\,\pm\,5$ yr} for the fragment pair of A and C of C/1882~R1 and
\mbox{$\Delta P = 284\,\pm\,7$ yr} for the pair of A and D.  Marsden's (1967) results showed
a slightly smaller differential momentum, equivalent to \mbox{$\Delta P = 177\,\pm\,5$ yr},
to apply to the orbital motions of the two fragments of C/1965~S1.  In order that perihelion
splitting of the parent sungrazer in an orbit with \mbox{$q = 1.09$ {\Rsun}} and \mbox{$P =
678$ yr} should release{\vspace{-0.06cm}} a fragment into an orbit with \mbox{$P = 922$ yr}, the
required separation velocity was 1.8~m~s$^{-1}$ in the direction of the orbital motion, comparable
to the separation velocities in Table~1 and by no means extreme.}  Marsden (1989) determined
that the ``original'' barycentric orbital period of C/1963~R1 was \mbox{$910\,\pm\,13$ yr}.
The bulk debris produced in the course of, and following, the event of 1041 became Population~Pe
of dwarf Kreutz sungrazers.  

Although C/1843 D1 and C/1882 R1 have been debated in the literature from just about every
possible angle, one issue has surprisingly been seldom addressed:\ Which of the two comets was
more spectacular?  Given that they arrived nearly 40-years apart and at different times of
the year (and therefore under different viewing conditions), their comparison is precarious.
Besides, there is the question of how exactly is the perceived level of performance measured
among comets?  By the overall impression of the object on the observer? By its apparent or
intrinsic brightness?  By the dimensions, such as the tail length, or by the tail prominence?
All criteria are questionable as they depend on the circumstances under which the comets are
judged, including the time of sighting, the observer's location, etc.
One would expect that comet observers and aficionados who had the opportunity to see both
sungrazers might have commented on their impressions (or be quoted by others to have done
so), but in one case that I followed the result was very disappointing.  Wlasuk's (1996)
comprehensive monograph on the life of Lewis Swift (1820--1913) describes that the Great
March Comet of 1843 ``dazzled'' him at the age of 23, but does not say a word about his
viewing the Great September Comet of 1882 at the age of 62!  A systematic search for more
positive evidence is beyond the scope if this investigation, but to illustrate the diversity
of opinion on the grandeur of the two sungrazers, I offer two examples.  On the one hand,
Lefroy (1882), regarding C/1882~R1 erroneously as the return of C/1843~D1, reported that
%
%
his friend, a witness to both celestial spectacles, had conveyed in a letter\footnote{Lefroy,
a~pioneer in the study of terrestrial magnetism, published an extract of the letter he
received from G.\ B.\ Bennett, his friend from Cape Town.  In the document dated nine days
after the perihelion passage of C/1882~R1, Mr.\ Bennett complained that, in reference to this
comet, ``\ldots Dr.~Gill is reported to have said, `the largest [comet] for 200~years.'  I
don't believe he said so; if so, he could not have seen the one of March, 1843.''  Mr.~Bennett
could not have been more correct, as 
Sir David Gill, Her Majesty's Astronomer at the Cape Observatory from 1879 to 1906, had been
born on 12~June~1843!  Gill's assistants at the observatory, Finlay \& Elkin (1882), were
more diplomatic than their boss, writing that ``\ldots (as far as can be gathered from the
accounts) [the comet of 1882] only resembles the one of 1843 in the point of extreme
brilliancy at perihelion.''}
that ``he does not consider the comet more conspicuous on this occasion than it was in 1843.''
On the other hand, Eddie (1883), a South African amateur astronomer, remarked on C/1882~R1
that it was ``the most brilliant comet of the \ldots century, not excepting the great comet
of 1843, which was \ldots not \ldots visible during the whole day.''\footnote{Eddie was born
two years after the appearance of C/1843~D1.}  More than a century later, two examples
illustrate that experts still have a problem to agree; while Moore (2000) maintains that
``[t]he brightest comet of modern times was probably that of 1843,'' it is C/1882~R1 that
in the verdict of Bortle (1998) was the ``[b]rightest, most extraordinary comet in over
1000~years.''  The results of my photometric study (Sekanina 2002) suggest that C/1843~D1 may
have been intrinsically fainter than C/1882~R1, but because of the large, poorly understood
systematic errors (integrated brightness vs brightness of nuclear condensation) of the
sporadic magnitude estimates available for either object, the former in particular, and the
uneven role of forward scattering effects, the issue should not be regarded as conclusively
resolved.  Comet C/1882~R1 was under observation much longer than C/1843~D1, but this
disparity was a consequence of the former comet's fragmentation at perihelion, which greatly
augmented the surface area available for activity. 
\pagebreak

The age of the Kreutz system, reckoned from the destruction of the contact-binary bond,
is estimated in this investigation at almost exactly two millennia, or 2$\frac{1}{2}$
revolutions about the Sun.  Although the scenario proposed here is clearly different from
the two-superfragment model contemplated by Sekanina \& Chodas (2004), the two age estimates
agree to within three centuries, or better than 15~percent.  They are much shorter than the
age estimates based on scenarios confining fragmentation events to perihelion.  Indeed,
Marsden (1967) estimated the age of the Kreutz system at 10--20~revolutions about the Sun,
or on the order of 10~millennia.  In the second paper he revised the estimate downward
(Marsden 1989) to a point that he, too, considered the comet of 372~BC as a possible
progenitor.  However, the time scale was compressed largely because most Kreutz sungrazers
were assigned unrealistically short orbital periods, as short as 360~years.  The many extra
sungrazers introduced make one ``painfully aware that the sequence \ldots requires that six
of the eight intermediaries that apparently passed perihelion unrecorded during the first
15~centuries of our era ought to have been truly spectacular objects'' (Marsden 1989).

\"{O}pik (1966), using similar but less elaborate arguments, estimated the Kreutz system's
age at 130~millennia, or more than 150~returns to perihelion.  The implications of such an
enormous age are daunting:\ even if each sungrazer should split at each perihelion into
only two (approximately equal) fragments, their total number after 150~returns would reach
10$^{45}$!  The progenitor's mass, even if as large as 10$^{20}$\,grams, would end up as
a population of micron-sized dust grains after some 100~returns, if the grains did not
sublimate away near perihelion in the meantime.  The net result is that after 150~returns
there would be no Kreutz system left.  A very short age is in fact inevitable, given the
recently observed massive sungrazers.

Should Aristotle's comet of 372 BC be the progenitor of the Kreutz system along the lines
proposed, it is likely that its appearance was more spectacular than that of either
C/1843~D1 or C/1882~R1, even though the level of increase in the extent of active surface
after the lobes decoupled is impossible to estimate.  It is also hard to judge whether the
chroniclers exaggerated the spectacle of the Aristotle comet and to what extent were its
purported credentials affected by the fact that it appeared during the exceptional period
of Ancient Greece.  One may even question the accuracy of Aristotle's account, given that
at the time the comet appeared he was a boy at the age of 11 or 12, but wrote the essay on
it in his book {\it Meteorologica\/} some 40~years later.     

The comet of 372 BC was a priori equally likely to be or not be part of the Kreutz system.
However, in the context of the proposed evolutionary path of the progenitor's Lobe~I, this
ambivalence about the comet's status appears to be strongly affected by the daylight comets
of AD~363 entering the timeline of the Kreutz system:\ in their roles of putative members,
the two episodes --- 372~BC and AD~363 --- do fit together as nicely as the final pair of
pieces in a puzzle.

Lastly, given the apparent propensity of Aristotle's comet to fragmentation that the
proposed scenario implies, there is no reason why the near-aphelion separation of the
lobes should have been the first fragmentation event the object underwent.  There is a
good chance that breakup episodes occurred near perihelion in 372~BC,\footnote{The popular
claim that Ephorus of Cyme observed the comet of 372~BC to split should not be taken
seriously, as a nuclear breakup can only be observed from a spacecraft positioned in the
comet's close proximity.  The terrestrial observer can at best detect an accompanying
outburst.  From the Earth, one is able to discern the nuclear condensation's duplicity (or
multiplicity), a breakup's product, not sooner than days, weeks, or months after the event,
and even this piece of evidence is strictly a telescopic phenomenon.  The Ephorus story,
repeated in the literature over and over again, is merely a myth, with no scientific merit
whatsoever.  This tall tale notwithstanding, it is entirely possible that --- unrecognized
--- the comet of 372~BC did indeed split at, or close to, perihelion.} and the
experience with C/1882~R1 suggests that fragments could have been scattered into orbits
whose periods were separated by a~century-or-so wide gaps.  In the adopted scenario the
main fragment of such a putative 372~BC event --- the contact-binary bond still intact
at the time --- ended up in an orbit with a period of about 735~yr to return, following
the near-aphelion breakup(s), in two or more pieces in late AD~363.  Another fragment
might have separated at that perihelion into an orbit with a period of 838~yr to return
as the comet of AD~467, yet another one into an orbit with a period of 623~yr to return
as the comet of AD~252, etc.  Additional breakups of both the parent and the fragments
themselves might have further augmented the number of such objects.  Strictly, the
fragments of this kind are not legitimate members of the Kreutz system, if its inception
is defined by the breakup of the contact-binary bond, yet they are of course related by
virtue of sharing the common parent.
 
\section{Conclusions}  
As knowledge of the Kreutz system has been advancing, the breadth of its recognized orbital
diversity has expanded from Marsden's (1967, 1989) original two-population (I, II) and
subsequent three-population (I, II, IIa) classification schemes to the currently proposed
eight independent populations (Pre-I, I, Ia, II, IIa, III, IIIa, IV) and a side branch
Pe of Population~I.  The major revision of the previous models of the Kreutz sungrazer
system was necessitated in order to accommodate new conceptual ideas, events, and a spate
of important data that only became available over the past decade or so.  The introduction
of Population~III was dictated by the appearance of C/2011~W3, while the rationale for
adding Population~Ia was more subtle:\ maintaining approximate uniformity in the stepwise
distribution of the populations' nodal longitudes over the interval of Populations~I--III.
Footprints of the new populations were retrieved by examining a carefully screened set of
Marsden's gravitational orbits for 193~dwarf Kreutz sungrazers imaged exclusively with the
C2 coronagraph on board the SOHO spacecraft.  Because of the ignored major effects of an
outgassing-driven acceleration in the motions of most dwarf Kreutz sungrazers, the computed
orbits defied the condition of shared apsidal line, valid for all naked-eye Kreutz comets.
In a plot of the nominal latitude of perihelion against the nominal longitude of the ascending
node for the set of select SOHO comets, the populations are discriminated from one another by
their nodal longitudes corrected for the out-of-orbit nongravitational effects.  In terms of
the true nodal longitude, adjacent populations are separated by gaps 9$^\circ$ to 10$^\circ$
wide on the average, while a total range of the nodal longitudes is 66$^\circ$.  Dwarf
comets of Populations~I, II, IIa, III, and Pe are associated with naked-eye sungrazers;
for Populations Pre-I, IIIa, and IV, they are yet to be discovered.  Their predicted
longitudes of the ascending node (for equinox J2000) are, respectively, near 11$^\circ$,
313$^\circ$, and 305$^\circ$, all outside the range of the five main populations.

With the image of the Kreutz system now fundamentally overhauled, the examination and
modeling of its populations are carried out primarily in terms of a sequence of early,
near-aphelion fragmentation events.  The initial episode is proposed to have involved a
breakup of the contact-binary progenitor into its two massive lobes, Lobe~I and Lobe~II,
and the low-mass connecting neck.  It is possible, if not likely, that the neck --- the
precursor of Population~Ia --- remained attached to one of the lobes (more probably Lobe~I)
over a period of time following the initial event.  In either case, the separation velocities
of the fragments, necessary to explain the large perturbations of the populations' mean
orbital elements, the longitude of the ascending node in particular, did not exceed
a few meters per second.  The most sizable surviving masses of Lobe~I and Lobe~II are
believed to be C/1843~D1 and C/1882~R1, respectively.  Lobe~II apparently was more than
Lobe~I susceptible to continuing fragmentation at very large heliocentric distance, near
aphelion.  Lobe~II was not only the parent to Population~II objects, but also the precursor
of the parent to Population~IIa objects; this parent (or its sibling with a lesser perihelion
distance) was the precursor of the parent to Population~III; which in turn was the precursor
of the parent to Population~IIIa; which was the precursor of the parent to Population~IV.
Lobe~I, the parent to Population~I objects, appears to have been fragmenting less profusely
on the kilometer-size scale, but probably more significantly on the dekameter-size scale,
judging from the dominance of Population~I among the SOHO sungrazers.  However, Lobe~I or
another precursor of C/1843~D1 or C/1963~R1 should have produced the primary fragment linked
to Population~Pre-I.  Comet C/1963~R1, a second-generation fragment of the precursor that
separated from Lobe~I in a fragmentation event following the progenitor's breakup, is
associated with Population~Pe, a side branch to Population~I.

The perceived relations among the products of cascading fragmentation, described by the
population classification scheme, are summarized in the pedigree chart.  Uncertainties grow
as one proceeds back in time, from the current generation of fragments (in the 19th--21st
centuries) to previous generations.  In the immediately preceding generation, the most likely
Kreutz system candidate was X/1106~C1, the parent to either C/1843~D1 or C/1882~R1.  Although
I distinctly prefer the former, either case can be argued.  The 12th century parent to one
of the two 19th century giant sungrazers was plainly missed, possibly because of extremely
unfavorable viewing geometry.  Fragments of the generation preceding the 12th century
generation are believed to have arrived at perihelion nearly simultaneously in the
4th~century AD; identities of possible candidates are offered, but only very incomplete
data are available.  An intriguing scenario involves comets seen in broad daylight in late
363, as recorded by the Roman historian Ammianus Marcellinus.  Although neither the number of
the comets nor their spread over time are known, inserting the year of this event as a third
link in the chain of returns in 1843 and 1106 is conducive to incorporating Aristotle's comet
of 372~BC --- the last perihelion passage of the contact-binary progenitor --- as a fourth,
and the earliest traceable, link.  Together they allude to a nearly constant orbital period
of 738~yr for the main residual mass of Lobe~I over an estimated age of the Kreutz system
of about two millennia, or 2$\frac{1}{2}$ revolutions about the Sun. Earlier perihelion
fragmentation of the comet of 372~BC, not involving the destruction of the contact-binary
bond, is by no means ruled out.

The proposed model differs dramatically from both the two-supefragment model (Sekanina \&
Chodas 2004) and the alternative model (Sekanina \& Chodas 2007), yet retaining some of
the features of either.  The new traits of the present model stem from the many significant
developments that have taken place since 2007.  As pointed out in Section~1, the arrival
of Comet Lovejoy (C/2011~W3) was perhaps the most important, but still only one, of these
events.  Irrevocably valid, and common to all three models, are the conclusions that by no
means have we seen the last spectacular Kreutz sungrazer and that another cluster of these
objects is on the way to perihelion, expected to arrive in the coming years and decades.
If anything, this case can now be argued more vigorously than ever before.    

In the interest of further progress in the understanding of the fragmentation history of
the Kreutz system, it is desirable that --- as the next milestone --- orbit integration
experiments be conducted, with full account of the planetary perturbations and simulated
nongravitational effects, in order to test and thereby confirm, refute, or impose limits on,
the proposed conceptual model.  Compared to the previous hypotheses, the current scenario is
straightforward and strongly rationale oriented; its ultimate success depends on whether, or
to what extent, it can be supported by rigorous computations.

{\small \bf Addendum.\/}  After completing this paper, I came across a brief remark by
Frisby (1883) that the orbital elements of the Great September Comet of 1882 {\it ``bear a
considerable resemblance to Comet I, B.C.\ 371; and it may possibly be its third return, a
very brilliant comet having been seen in full daylight A.D.~363.''\/}  Apart from the fact
that the orbit{\vspace{-0.04cm}} of the comet of 372~BC (or $-$371, not 371~BC) is not
known,\footnote{The set of elements published by Pingr\'e (1783; see also Kronk 1999),
which does not at all bear a resemblance to that of the Great September Comet of 1882
(or any other Kreutz sungrazer for that matter) in terms of spatial orientation, is
merely a wild guess; it implies a perihelion longitude of about 180$^\circ$ (equinox
J2000), compared to the Kreutz system's 282$^\circ$.}
this statement is --- to my knowledge --- one of only two instances that suggest a possible
link of the spectacle of AD~363 to the Kreutz system, the other being the noted narrative
by Seargent (2009), commented on in Section~5.3.  The reference to ``full daylight'' makes
it clear that Frisby meant the Roman, not the Chinese, event.  It is remarkable that,
relying on his own orbit computations, Frisby had apparently no qualms whatsoever about
the orbital period of the Great September Comet of 1882 in a range of 750--800~yr at the
time when Kreutz's (1883) value was 843~yr, Elkin's (1883) best guess hovered near 1500~yr,
and not long after Chandler (1882a, 1882b) had derived a period of 4000~yr.  It was not
until five years later that Kreutz (1888) considered the potential identity of the 1882
comet with that of 1106 and it took him another three years to arrive at his final verdict
on the pre-fragmentation orbital period of the 1882 comet by estimating its lower limit at
770~yr and upper limit at 1000~yr (Kreutz 1891).\\[0.1cm]

This research was carried out at the Jet Propulsion Laboratory, California Institute of
Technology, under contract with the National Aeronautics and Space Ad\-min\-istration.\\

\begin{center}
{\footnotesize REFERENCES}
\end{center}
\vspace{-0.4cm}
\begin{description}
{\footnotesize
\item[\hspace{-0.3cm}]
Barrett, A. A. 1978, J. Roy. Astron. Soc. Canada, 72, 81
\\[-0.57cm]
\item[\hspace{-0.3cm}]
Bortle, J. E. 1998, {\tt http://www.icq.eps.harvard.edu/bortle.html}
\\[-0.57cm]
\item[\hspace{-0.3cm}]
Chandler, S. C., Jr. 1882a, Nature, 27, 81 
\\[-0.57cm]
\item[\hspace{-0.3cm}]
Chandler, S. C., Jr. 1882b, Astron. Nachr., 103, 347 
\\[-0.57cm]
%
%
%
\item[\hspace{-0.3cm}]
Eddie, L. A. 1883, Mon. Not. Roy. Astron. Soc., 43, 289
\\[-0.57cm]
\item[\hspace{-0.3cm}]
Elkin, W. L. 1883, Astron. Nachr., 104, 281 
\\[-0.57cm]
\item[\hspace{-0.3cm}]
Elton, H. 2018, The Roman Empire in Late Antiquity:\ A Political{\linebreak}
 {\hspace*{-0.6cm}}and Military History.  Cambridge, UK: Cambridge University{\linebreak}
 {\hspace*{-0.6cm}}Press, 378pp 
\\[-0.57cm]
\item[\hspace{-0.3cm}]
England, K. J. 2002, J. Brit. Astron. Assoc., 112, 13
\\[-0.57cm]
\item[\hspace{-0.3cm}]
Finlay, W. H., \& Elkin, W. L. 1882, Mon. Not. Roy. Astron. Soc.,{\linebreak}
 {\hspace*{-0.6cm}}43, 24
\\[-0.57cm]
\item[\hspace{-0.3cm}]
Frisby, E. 1883, Astron. Nachr., 104, 159
\\[-0.57cm]
\item[\hspace{-0.3cm}]
Hasegawa, I.\,1979, Publ.\,Astron.\,Soc.\,Japan,\,31,\,257\,(errata\,p.\,829)
\\[-0.57cm]
\item[\hspace{-0.3cm}]
Hasegawa, I., \& Nakano, S.\,2001,\,Publ.\,Astron.\,Soc.\,Japan,\,53,\,931
\\[-0.57cm]
%
%
\item[\hspace{-0.3cm}]
Ho, P.-Y. 1962, Vistas Astron., 5, 127
\\[-0.57cm]
\item[\hspace{-0.3cm}]
Kreutz, H. 1883, Astron. Nachr., 104, 157 
\\[-0.57cm]
\item[\hspace{-0.3cm}]
Kreutz, H. 1888, Publ. Sternw. Kiel, 3
\\[-0.57cm]
\item[\hspace{-0.3cm}]
Kreutz, H. 1891, Publ. Sternw. Kiel, 6
\\[-0.57cm]
\item[\hspace{-0.3cm}]
Kreutz, H. 1895, Astron. Nachr., 139, 113
\\[-0.57cm]
\item[\hspace{-0.3cm}]
Kreutz, H. 1901, Astron. Abh., 1, 1
\\[0.57cm]

\item[\hspace{-0.3cm}]
Kronk, G. W. 1999, Cometography: A Catalogue of Comets.{\linebreak}
 {\hspace*{-0.6cm}}I. Ancient--1799. Cambridge, UK: Cambridge University Press,{\linebreak}
 {\hspace*{-0.6cm}}580pp
\\[-0.57cm]
\item[\hspace{-0.3cm}]
Lefroy, J. H. 1882, Nature, 26, 623
\\[-0.57cm]
%
\item[\hspace{-0.3cm}]
Lenski, N. 2002, Failure of Empire:\ Valens and the Roman State{\linebreak}
 {\hspace*{-0.6cm}}in the Fourth Century A.D.  Berkeley, CA: University of Cali-{\linebreak}
 {\hspace*{-0.6cm}}fornia Press, 454pp 
\\[-0.57cm]
\item[\hspace{-0.3cm}]
Marsden, B. G. 1967, AJ, 72, 1170
\\[-0.57cm]
\item[\hspace{-0.3cm}]
Marsden, B. G. 1989, AJ, 98, 2306
\\[-0.57cm]
\item[\hspace{-0.3cm}]
Marsden, B. G. 2005, Annu. Rev. Astron. Astrophys., 43, 75
\\[-0.57cm]
\item[\hspace{-0.3cm}]
Marsden, B. G., \& Roemer, E. 1978, Quart. J. Roy. Astron. Soc.,{\linebreak}
 {\hspace*{-0.6cm}}19, 38
\\[-0.57cm]
\item[\hspace{-0.3cm}]
Marsden, B. G., \& Williams, G. V. 2008, Catalogue of Cometary{\linebreak}
 {\hspace*{-0.6cm}}Orbits 2008, 17th ed. Cambridge, MA:\ Minor Planet Center/{\linebreak}
 {\hspace*{-0.6cm}}Central Bureau for Astronomical Telegrams, 195pp
\\[-0.57cm]
\item[\hspace{-0.3cm}]
Marsden, B.\,G., Sekanina, Z., \& Yeomans, D.\,K.\,1973,\,AJ,\,78,\,211
\\[-0.57cm]
%
%
\item[\hspace{-0.3cm}]
Moore, P. 2000, The Data Book of Astronomy.\ Bristol:\,Institute\,of{\linebreak}
 {\hspace*{-0.6cm}}Physics Publishing, p.\,233
\\[-0.57cm]
%
%
\item[\hspace{-0.3cm}]
\"{O}pik, E. J. 1966, Ir. Astron. J., 7, 141
\\[-0.57cm]
\item[\hspace{-0.3cm}]
Pingr\'e, A. G. 1783, Com\'etographie ou Trait\'e Historique et{\linebreak}
 {\hspace*{-0.6cm}}Th\'eorique des Com\`etes.  Paris:\ L'Imprimerie Royale
\\[-0.57cm]
\item[\hspace{-0.3cm}]
Ramsey, J. T. 2007, J. Hist. Astron., 38, 175
\\[-0.57cm]
\item[\hspace{-0.3cm}]
Rolfe, J. C. 1939, The Roman History of Ammianus Marcellinus,{\linebreak}
 {\hspace*{-0.6cm}}Book XXX.\ {\tt https://penelope.uchicago.edu/Thayer/E/Roman/}{\linebreak}
 {\hspace*{-0.6cm}}{\tt Texts/Ammian/30$^\ast$.html}
\\[-0.57cm]
\item[\hspace{-0.3cm}]
Rolfe, J, C. 1940, The Roman History of Ammianus Marcellinus,{\linebreak}
 {\hspace*{-0.6cm}}Book XXV.\ {\tt https://penelope.uchicago.edu/Thayer/E/Roman/}{\linebreak}
 {\hspace*{-0.6cm}}{\tt Texts/Ammian/25$^\ast$.html}
\\[-0.57cm]
\item[\hspace{-0.3cm}]
Seargent, D.\,2009, The Greatest Comets in History:\ Broom Stars{\linebreak}
 {\hspace*{-0.6cm}}and Celestial Scimitars. New York:\ Springer Science+Business{\linebreak}
 {\hspace*{-0.6cm}}Media, LLC, 260pp
\\[-0.57cm] 
\item[\hspace{-0.3cm}]
Sekanina, Z. 2002, ApJ, 566, 577
\\[-0.57cm]
\item[\hspace{-0.3cm}]
Sekanina, Z., \& Chodas, P. W. 2002, ApJ, 581, 760
\\[-0.57cm]
\item[\hspace{-0.3cm}]
Sekanina, Z., \& Chodas, P. W. 2004, ApJ, 607, 620
\\[-0.57cm]
\item[\hspace{-0.3cm}]
Sekanina, Z., \& Chodas, P. W. 2007, ApJ, 663, 657
\\[-0.57cm]
\item[\hspace{-0.3cm}]
Sekanina, Z., \& Chodas, P. W. 2012, ApJ, 757, 127
\\[-0.57cm]
\item[\hspace{-0.3cm}]
Sekanina, Z., \& Kracht, R. 2015, ApJ, 801, 135
\\[-0.57cm]
\item[\hspace{-0.3cm}]
Sierks, H., Barbieri, C., Lamy, P. L., et al.\ 2015, Science, 347, a1044
\\[-0.57cm]
\item[\hspace{-0.3cm}]
Stern, S. A., Weaver, H. A., Spencer, J. R., et al.\ 2019, Science,{\linebreak}
 {\hspace*{-0.6cm}}364, 9771
\\[-0.57cm]
\item[\hspace{-0.3cm}]
Strom, R. 2002, A\&A, 387, L17
\\[-0.62cm]
\item[\hspace{-0.3cm}]
Wlasuk, P. T. 1996, Quart. J. Roy. Astron. Soc., 37, 683}
%
\end{description}
\vspace{0.47cm}
\end{document}